\documentclass[12pt,a4paper]{article}
\usepackage{eurosym}
\usepackage[a4paper,top=3cm,bottom=3cm,left=3cm,right=3cm]{geometry}
\usepackage{algorithm2e,amsfonts,amsmath,mathtools,amsthm,amssymb,authblk}  
\usepackage{cite}
\usepackage{color}
\usepackage[footnotesize]{caption}
\usepackage[english]{babel}
\usepackage{enumerate,enumitem}
\usepackage{epigraph}
\usepackage{float}
\usepackage{graphics,graphicx,latexsym,amsfonts}
\usepackage{lineno}
\usepackage{multirow}
\usepackage{multicol}
\usepackage{colortbl}
\usepackage{pst-plot,pstricks}
\usepackage{picture}
\usepackage{subfig}
\usepackage{url}
\usepackage{hyperref}
\usepackage[symbol]{footmisc}

\providecommand{\U}[1]{\protect \rule{.1in}{.1in}}
\usepackage{setspace}
\onehalfspace 
\title{Authoritarianism vs. democracy:
\\ Simulating responses to disease outbreaks}
\author{
A.E. Biondo\footnote{Dept. of Economics and Business, University of Catania, \textit{corresponding}: ae.biondo@unict.it},
	 G. Brosio\footnote{Dept. of Economics and Statistics ``Cognetti De Martiis'', University of Turin}, 
	 A. Pluchino\footnote{Dept. of Physics and Astronomy, University of Catania and INFN - Section of Catania}, 
     R. Zanola\footnote{Dept. of Law and Political, Economic, and Social Sciences,  University of Eastern Piedmont.}
}

\def\hk{\textsc{hk}}
\def\hkk{\textsc{hk }}

\def\hkkp{\textsc{hk-p }}
\date{ }

\begin{document}

\maketitle

\begin{abstract}
Disease outbreaks force the governments to rapid decisions to deal with. However, the rapid stream of decision-making could be costly in terms of the democratic representativeness. The aim of the paper is to investigate the trade-off between pluralism of preferences and the time required to approach a decision. To this aim we develop and test a modified version of the Hegselmann and Krause (2002) model to capture these two characteristics of the decisional process in different institutional contexts. Using a twofold geometrical institutional setting, we simulate the impact of disease outbreaks to check whether countries exhibits idiosyncratic effects, depending on their institutional frameworks. Main findings show that the degree of pluralism in political decisions is not necessarily associated with worse performances in managing emergencies, provided that the political debate is mature enough. \\
	
\textbf{JEL} Classification: C53, C63, H11, H12.
\\


\end{abstract}

\section{Introduction}

The World Health Organization (\textsc{who}) has declared the 2020 pandemic of coronavirus named  \textsc{covid}-19 a global public health emergency. In less than seven months the virus has spread from a limited number of epicentres in Asia (mainly China and Korea) to practically every country. Uncertainty about the future of pandemic is growing (immunity, intensity of contagion, future vaccines, etc.) and it nourishes speculation on national responses and feeds a worldwide debate on the best form of government to contrast it. 

In the early months of the pandemic China locked down millions of people who lived in an area with major outbreaks, with invasive surveillance and coercion. In Hungary, Prime Minister was allowed by the parliament to rule without any control and time limit. Analogously, the autocratic regime of Venezuela faced with limited contagion, but with a health sector in total disarray adopted immediately strict lockdown measures. The opposition interpreted them as a political strategy to cut the manifestation of growing dissent against the regime. Belarus, the most autocratic regime in Europe, delayed a response. As of April 23 restaurants, coffee shops and movie theatres remain open. Professional soccer was in full swing. In the capital, Minsk, the subways was crowded. Most businesses required workers to show up. As of April 27, there were 10,463 cases and 72 deaths. A state of emergency was introduced in Kyrgyzstan, a Central Asia authoritarian regime, although the country had been spared from the pandemic. As of April, 695 cases were recorded and only 8 deaths. Other states in the region, such as Turkmenistan, one of the most isolated states in the world, held a series of mass sporting events despite. According to official sources, it had no case of coronavirus infection as of April 27. 

By contrast, since time and information to solve an issue involving the diversity of constituents in collaborative co-deciding is costly, democratically elected leaders in Italy, Spain, UK, United States were criticized for delaying adopting stringent measures to coherently contain the pandemic. An exception being India, the world's largest democracy, which applied heavy-handed measures in defeating the coronavirus. Germany claimed and was credited with taking early response, although without strict containment measures. This did not contain the spread of the virus and it confronted with high fatality rates.

A key issue the debate is going through is whether authoritarian countries perform better than democratic ones to contrast pandemics (Alon et al. 2020). For instance, although the mix of strategies (and effects) to respond to the coronavirus, Kleinfeld (2020) sustains the superiority of the democratic model. Berengaut (2020) affirms that public health depends on public trust as it is the case of Western nations, irrespective of influential admirers of the Chinese decisive intervention. Schmemann (2020) questions the cures for  \textsc{covid}-19, while Bieber (2020) highlights the risk of abuse for both dictatorships and democracies. The uncertainty about which government is better to answer to coronavirus is discussed in Khavanag (2020), who suggests that Chinese autocratic rules seem to prove to be successful, but, at the same time, he also suggests that open media which guarantee democratic politics are time-consuming more than inadequacy. This present debate on pandemic response has plenty of antecedents in political literature. The theory of the political survival of government affirms that democratic governments are prone to take quick actions to help victims of disasters to maximize their time in office (Bueno de Mesquita et al., 2003; Morrow et al., 2008). Moving the logic of political survival to the 2009 H1N1 pandemic, Baekkeskov and Rubin (2014) show that mature democracies adopted massive vaccination as a precautionary strategy the cost of which is much lower than the cost of failing subsequent election. Moreover, the response to pandemic could be strictly correlated with the country electoral cycle, calling for more intensive actions the shorter the period to the next election. Finally, the response intensity depends upon the degree and intensity of political participation which is expected to vary across democracies. In summary, according to Baekkeskov and Rubin, it is not democracy per se which can suggest the response to the pandemic, but how democracy operates de facto. A clear distinction between democratic vs. autocratic response to pandemics emerges in Schwartz (2012) who compares China's and Taiwan's response to the SARS outbreak in 2002/3. Why, the author wonders, did authoritarian China succeed while democratic Taiwan failed to respond to SARS? To explore the issue Schwartz makes a useful distinction between routine crises and novel crises. In the first case, politicians defer to operational commanders moderate responses, usually experienced in similar crises. By contrast, novel crises, being pandemic an example of, are much more insidious, due to the delay in recognizing the true nature of them and, consequently, to set adequate responses. This delay to identify priorities and actions is strengthened in democratic countries, where novel emergencies imply multiple jurisdictions and different levels of representative government to coordinate to deal with difficulties. They recognize two different delays in contrasting actions. Firstly, the lack of disease surveillance to detect outbreaks. Secondly, the delay to recognize the international concern of the outbreaks, a case in which political mobilization plays a fundamental role. The distinction between democratic and autocratic regimes is here irrelevant to explain different mobilization. Rather, disease severity, globalization, disease spread to the whole population rather than smaller groups, its perception by the public, are key factors to explain countries' response. This is not the position expressed in Burkle (2020). Autocratic regimes are incapable of understanding the health consequences on populations of pandemics causing sensible delay to manage crises, placing the rest of the world at increasing risk.

In summary, analysing the present debate on governments? responses to pandemic, it seems to emerge a trade-off between the democratic representativeness of opinions and the necessity to be quick to deal with outbreaks. Dictatorships are, in principle, faster than democracy by neglecting the time-demanding coordination effort, but at the cost of ignoring the consensus of the entire population. The aim of the paper is to investigate the trade-off between time and representativeness of preferences, simulating the performance of different political systems facing an emergency, being pandemic a clear example of an emergency problem involving an entire community. 

Models of opinion dynamics are particularly helpful in providing the minimal set of analytical tools describing the \textit{consensus formation}. They have received a wide interest, from different contexts, e.g., from mathematics, physics, and economics. Early examples of such contributions are Harary (1959), French (1956), Abelson (1967) and DeGroot (1974), among others. More recent models are Lehrer and Wagner (1991) and Friedkin and Johnsen (1999) and, with regards to non linear models in Krause (1997 and 2000), Hegselmann and Flache (1998), Deffuant \textit{et al.} (2000), and Sznajd-Weron and Sznajd (2000). Acemoglu and Ozdaglar and Lorenz (2007) interestingly survey relevant parts of the topic. Many opinion dynamics models can be considered, indeed, as a class of agent-based models (ABMs). Such frameworks, in which the interaction among different individual entities is able to generate emerging aggregate outcomes, are particularly useful when dealing with complex evolving socio-economic systems, for their ability in characterizing micro motives within macro dynamics (Tesfatsion, 2006a, 2006b). By means of the bottom-up design of such a class of models, the analysis of complexity and emergent phenomena is in fact possible (Delli Gatti \textit{et al.}, 2008, 2011, and Fagiolo and Roventini, 2008). The early origin of ABMs can be traced back to the von Neumann self-reproducing cellular automata in the first half of last century (von Neumann and Burks, 1966), but their full expansion has started in the 90s, thanks to the sudden increase of computational power (Epstein and Axtell, 1996, Bonabeau, 2002). Particularly interesting have been the further developments of ABMs in the context of sociophysics (as in Galam, 2002) and econophysics, where they have been applied to the analysis of financial markets (Mantegna and Stanley 1999, Chakraborti \textit{et al.} 2011a and 2011b) and adopted to general macroeconomic modelling, as surveyed in Dawid and Delli Gatti (2018). Their relevance in other branches of economics has been rapidly growing in the recent past and is increasingly popular, due to the combination of their versatility in theoretical design and their privileged feature of being able to replicate several stylized facts of empirical data. Examples are, among others, Chang (2011) and (2015) for industrial organization, Neugart and Richiardi (2012) for labour economics, Djvadian and Chow (2017) for transport economics, Kirman (2011) for behavioral economics, Adami et al. (2016), for game theory. 

To this aim, we develop and test a modified version of the Hegselmann and Krause (2002), the \hk-Politics, \hkkp henceforth. Two different characteristics are considered, namely, the democratic representativeness and the time efficiency. Correspondingly, the sequential policy-making problem is described in two steps: firstly, the political debate showing the degree of free circulation of ideas and the possibility of discussion among different points of view; secondly, the converging process leading to the final policy decision. Using a two-fold geometrical institutional setting, referred to specific indicators, i.e.  the press freedom and the democracy index, a representative sample of twenty countries is adopted to build an original index to capture the distortion in the political representation of citizens' preferences relative to the efficiency of decisional time requirements. Information systems reveal to be crucial in the diffusion of values and priorities in public planning problems (Hoos, 1971), where the strategic weight of decisions is rooted in the ability of the decision-maker to manage the situation at hand, analogously to what happens in managerial decisions (Green and Kolesar, 2004). In our benchmark model, we simulate the impact of disease outbreaks to check whether countries exhibit idiosyncratic effects, depending on their institutional frameworks. Our main findings show that the degree of pluralism in political decisions is not necessarily associated with worse performances in managing emergencies, provided that the political debate is mature enough. Thus, even in case of emergencies, the concentration of power does not necessarily lead to desired improvements in the efficiency of policy actions.

The remainder of the paper is organized as follows: section two presents the model; section three illustrates the case study with simulation results; section four contains concluding remarks.

\section{The HK-P model}

In a model of opinion dynamics, agents are typically described by means of their opinion profiles, consisting in vectors of length $n\ge1$, according to the dimension of the preference space in which they are modeled. Thus, a society can be described as a community of $N$ individuals with different opinions, influencing each other. The decisional process is represented by the route to convergence towards one or more non-reducible states. Such states are \textit{steady states}, for the dynamic forces driving the adaptation rest in that condition. 

As typical in bounded confidence opinion dynamics models, we will restrict our attention to the case of a partial interaction of agents, deriving from the spatial proximity of their opinions. This appears quite realistic, since political mediation is reasonably possible even between far, but still not incompatible, positions. Proximity is, then, proposed as a measure of compatibility: each agent will consider only opinions within a given neighbourhood of the opinion space around himself, i.e., a compatibility interval. The opinion space $\mathbf{\Omega}=\left[0,1\right]^n$ is an hypercube with $n$ dimensions and open boundary conditions, whose points define profiles of each individual $i$ as $n$-dimensional vectors of the type $\mathbf{x}_i=\left[x_{i, 1}, x_{i, 2}, ..., x_{i, n}\right]$, for $i=1, 2, ..., N$, collecting the opinions about each of the $n$ political topics. Consistently with the number of dimensions, the compatibility interval is represented by an hypersphere with radius of length $\varepsilon_i=\varepsilon$, being $\varepsilon \in [0,1]$ the confidence bound. Thus, the compatibility interval of each agent can be defined as $B(\mathbf{x}_{i},\varepsilon)$, $\forall i$. The main dynamic feature of the \hkk model consists in the update of each opinion vector $\mathbf{x}_i(t)$ which, at time $t+dt$, becomes equal to the average of all and only opinion vectors included within $B(\mathbf{x}_{i},\varepsilon)$. 

Such a dynamics produces different outcomes according to the value of the confidence bound. Below a critical threshold $\varepsilon_c$, it asymptotically generates a non-reducible state with clusters of opinions, $n_\chi \in [2, N]$, such that $\lim_{\varepsilon \rightarrow 0} n_\chi = N$ and $\lim_{\varepsilon \rightarrow \varepsilon_c} n_\chi = 2$. Above such a critical threshold, consensus is always achieved, i.e., $n_\chi =1$, independently of the initial opinions distribution. 
In Fortunato (2005a), it has been shown that the value of $\varepsilon_c$ strictly depends on the type of graphs which models the community of the $N$ interacting agents. For a complete graph, i.e. for a fully interacting community, one finds that $\varepsilon_c=0.2$. 

In the limit of an infinite number of agents it is convenient to assume a continuous time version of the \hkk dynamics, defining a uniform opinion distribution function $P(\mathbf{x}, t)$ such that

\begin{equation}
\int_\Omega P(\mathbf{x}, t)d\mathbf{x} = N(t)
\label{tot-pop}
\end{equation} 

\noindent is the count of the entire population at any $t$. Following some of the advances proposed in Fortunato \textit{et al.} (2005b), the rate equation of the model is described by the following integro-differential equation:

\begin{equation}
\frac{\partial}{\partial t}P(\mathbf{x},t)=
\int_\Omega P(\mathbf{x}_1,t)
\delta\Big(\mathbf{x}-\bar{\mathbf{x}}_{B(\mathbf{x}_{1},\varepsilon)} \Big) d\mathbf{x}_1
- \int_\Omega P(\mathbf{x}_1,t)
\delta\Big(\mathbf{x}-\mathbf{x}_1\Big) \, 
 d\mathbf{x}_1
\label{opinion-dynamics}
\end{equation}

\noindent where $$\bar{\mathbf{x}}_{B(\mathbf{x}_{1},\varepsilon)} \equiv \frac{\int_{B(\mathbf{x}_{1},\varepsilon)}\mathbf{x}_{0}\, P(\mathbf{x}_{0},t) \, d\mathbf{x}_{0}} 
{\int_{B(\mathbf{x}_{1},\varepsilon)}P(\mathbf{x}_{0},t)\, d\mathbf{x}_{0}}$$ is the average opinion profile calculated over the hypersphere $B(\mathbf{x}_{1},\varepsilon) \subset \Omega$, centered at $\mathbf{x}_{1}$ with diameter $\varepsilon$, and $\delta(\mathbf{x})=\prod_j\,\delta (x_j)$, with $j=1, 2, ..., n$, being $\delta(\cdot)$ the Dirac delta function. Eq.(\ref{opinion-dynamics}) captures the \hkk dynamics explained above: indeed, the variation of $P(\mathbf{x},t)$ at each $\mathbf{x}$, in the time interval $dt$, is the net sum of two components: the first one, constituted by the positive contribution of all incoming opinion vectors $\mathbf{x}_1$, when $\bar{\mathbf{x}}_{B(\mathbf{x}_{1},\varepsilon)} = \mathbf{x}$; the second one, constituted by the negative contribution of the outgoing opinion vectors $\mathbf{x}_1$, when $\mathbf{x}_1=\mathbf{x}$. 

In this study, dealing with a finite number of agents, $N$, we build our \hkkp model by considering a discrete time version of the \hkk dynamics on a complete graph. We refer to a $2D$ opinion space, whose dimensions can be interpreted as the set of policy tools ($x$ axis) and policy targets ($y$ axis). In this space, the placement of opinions should not be intended as the traditional perspective of left- or right-party positions. Instead, it stands for the variety of ideas in terms of specific objectives and methodologies to adopt in order to obtain desired results. Each one of our $N$ agents, or politicians, is endowed with an individual political profile, i.e. the opinion $\mathbf{x}_i=[x_{i,1}, x_{i,2}]$, $i=1, 2, ..., N$, which will be defined as a point in the two-dimensional opinion space. Thus, we define the compatibility interval $B(\mathbf{x}_i,{\varepsilon})$ as a circle of radius equal to the confidence bound $\varepsilon$. 

At $t=0$, the $N$ political profiles are uniformly distributed in the allowed opinion space. At each subsequent time step, agents adopt a parallel update process in order to adapt their opinions in response to compatible ones, according to the following update rule:

\begin{equation}
\mathbf{x}_i(t+1)=\frac{\sum_{j:\|\mathbf{x}_i(t)-\mathbf{x}_j(t)\|<\varepsilon} a_{ij} \mathbf{x}_j(t)}{\sum_{j:\|\mathbf{x}_i(t)-\mathbf{x}_j(t)\|<\varepsilon} a_{ij}}
\label{discrete}
\end{equation}

\noindent where $\|\mathbf{x}_i(t)-\mathbf{x}_j(t)\|$ is the metric distance between the opinion vectors $i$-th and $j$-th, and $a_{ij}$ is the adjacency matrix of the graph. Since we consider here a complete graph, it is $a_{ij}=1$ for $i \neq j$ and $a_{ij}=0$ for $i=j$. According to equation (3), at time $t+1$, $\mathbf{x}_i(t+1)$ will become equal to the average of all the opinion vectors included, at time $t$, within the compatibility circle of agent $i$-th.

In order to address the dynamics of a set of $K$ countries, with different political systems, in the \hkkp model we will relate the size of the $2$D opinion space and the confidence bound $\varepsilon$ with two main attributes of the political debate within a given country $C_k$, namely, the press freedom and the democracy index. The former, quantified by the parameter $b_{\text{\textsc{pf}}}$, will be related to the initial setup of the opinion space, as it expresses the attitude of the policy-maker to restrict, by means of subservient media, the official opinion space in order to control the focus of the political debate. We will underline later that the sudden occurrence of an emergency situation could be translated in a further reduction of the allowed space, proportional to the level of alarm. The latter, quantified by the democracy index $b_{\text{\textsc{d}}}$, will refer to the attitude of individuals to discuss their opinions and approach a partial or a total common consensus, taking into account the decentralized points of view allowed by the level of democracy in the country. This process will be regulated by the value of the confidence bound, tuned according to the democracy index, which should be able to distinguish in a natural way democratic systems from authoritarian ones. We provide below, in a dedicated section, the description of both used indexes and their wide adoption in literature.

All in all, the model here proposed is aimed to describe the process of convergence to a final decision, starting from initial different positions, given the institutional features of a country. 
Thus, each country $k$ of the considered set will be characterized by the couple $C_k= \left\{ b^k_{\text{\textsc{pf}}}, b^k_{\text{\textsc{d}}} \right\}$, $k=1,...,K$, which defines its typical configuration for the modelled dynamics, as explained in next sections.

\begin{figure}[h]
	\begin{center}
		\includegraphics[width=5in,angle=0]{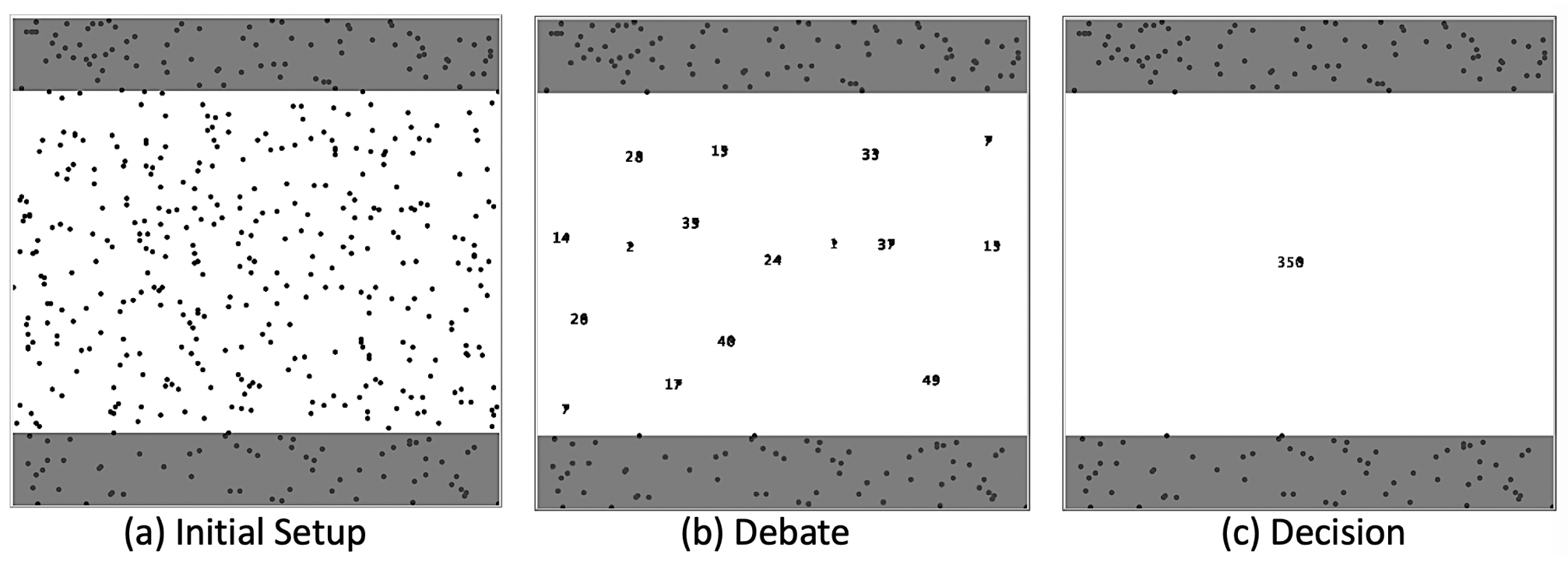}
		\caption{\small 
			(a) An example of initial setup for the simulation of a generic country $C_k$, with non complete press freedom, $b^k_{\text{\textsc{pf}}}=0.7$, and absence of any emergency pressure, $e=0$. At $t=0$, the $N=500$ individual opinions (or agents, represented by red dots) are uniformly distributed over the whole space. The limited press freedom in the country reduces the allowed space (coloured in white) for the subsequent opinion dynamics, thus censoring part of them. Opinions (agents) outside from the allowed space represents extreme (or simply uncomfortable) positions that remain off from the official debate. \\ (b) If in the considered country $C_k$ is a democratic system, i.e., $b^k_{\text{\textsc{d}}} > b^{th}_{\text{\textsc{d}}}$, a free political debate is permitted. Thus,  the $N^k_{\text{\textsc{pf}}} = 350$ opinions involved in the dynamics approach a situation in which political agreements are reached. In other words, the \hkkp dynamics leads to a fragmented state with several clusters (each labelled by its size). \\ (c) After political clusters have been formed, such $N^k_{\text{\textsc{pf}}}$ opinions are finally collapsed over just a single point, representing the final decision. In an authoritarian country (i.e. a country with $b^k_{\text{\textsc{d}}} < b^{th}_{\text{\textsc{d}}}$), the political debate (b) is not even experienced, since the dynamics drives directly towards the final decision (assumed by either the dictator or the regime).              
		}
		\label{initial-setup} 
	\end{center}
\end{figure}

\subsection{Initial Setup ($t=0$): Press freedom and emergency pressure}

The press freedom index $b_{\text{\textsc{pf}}}$ determines the boundaries of the opinion space of the \hkkp model. The simulated opinion space is a square world of $300 \times 300$ units where, as above explained, the two dimensions represent, respectively, the variety of targets and tools discussed in the policy debate. The width of the policy targets (the vertical axis) is influenced by the press freedom, in terms of reduction of the opinion space due to a specific direction operated by the media communication. 

For a given country $C_k$ with press freedom index $b^k_{\text{\textsc{pf}}}$ the reduction, applied before starting any simulation, is modelled by restricting the opinion space as $300 \times (300 \cdot b^k_{\text{\textsc{pf}}})$, with $0 <  b^k_{\text{\textsc{pf}}}\le1$. Then, a country with complete press freedom ($b^k_{\text{\textsc{pf}}}=1$) will be characterized by the full opinion space, whereas, a country with reduced press freedom will be depicted by a correspondingly reduced space. In Figure 1(a), an example of initial conditions in the opinion space of a country is shown. The press freedom is non complete, i.e.,$b^k_{\text{\textsc{pf}}}=0.7$ and $N=500$ opinion profiles $\mathbf{x}_i(0), i=1,...,N$, represented by red dots, are uniformly distributed at random over the whole space. However, the subsequent dynamics - which will be addressed in the next section - will take into account only the reduced number of opinions/agents included in the allowed space (coloured in white). For $b^k_{\text{\textsc{pf}}}<1$, this number will be lower than $N$, let us call it $N^k_{\text{\textsc{pf}}}$. In general, for $N \gg1$, one has that $N^k_{\text{\textsc{pf}}}$ tends to $\bar{N}^k_{\text{\textsc{pf}}} \sim N \cdot b^k_{\text{\textsc{pf}}}$. 

The appearance of an emergency can worsen the press freedom situation by adding a second restriction to the allowed opinion space. In particular, an emergency parameter $e\ge0$, equal for all the countries, can be introduced in order to further reduce the press freedom parameter, according to the following rule: $b^k_{\text{\textsc{pfe}}} = b^k_{\text{\textsc{pf}}} \cdot (100 - e) / 100 $. In this situation we will have only $N^k_{\text{\textsc{pfe}}} \sim N \cdot b^k_{\text{\textsc{pfe}}}$ opinions/agents involved in the dynamics. 

Summarizing, we interpret the reduction of the opinion space as the censoring activity operated either surreptitiously by the regime (controlling the press freedom) or spontaneously by the mass-media (with regards to the emergency situation). Our model will present a possible measure for the negative effects of such a censoring activity and the consequent loss of democratic representativeness due to the exclusion of all opinions out from tolerated positions.    

\subsection{Dynamics ($t>0$): Democracy index, debate and decision}

The second main parameter of the \hkkp model is the democracy index $b_{\text{\textsc{d}}}$, which regulates the dynamics for $t>0$. We assume that $0 <  b_{\text{\textsc{d}}} < 1$, being $1$ the ideal value for a perfect democracy and $0$ the ideal value of a perfect dictatorship. In democratic systems, one expects to have a political debate before the convergence towards the final consensus is reached. In other words, the free discussion leads to a dynamical fragmentation of opinions in several (more than one) clusters, representing a plurality of different points of view about both policy objectives and tools required to obtain them. On the other hand, for authoritarian systems, one can expect that any decision will be taken without any intermediate debate, thus the dynamical process should directly produce a single cluster in the opinion space, representing the achievement of the final consensus. These two qualitatively different behaviours can be obtained by putting in a one-to-one correspondence the democracy index with the confidence bound $\varepsilon$, which will become a function $\varepsilon=f(b_{\text{\textsc{d}}})$ as explained below.

Let us imagine to perform a ranking of the set of $K$ countries, ordered according to decreasing values of their democracy indexes $b^k_{\text{\textsc{d}}}$, so that $b^k_{\text{\textsc{d}}}>b^{k+1}_{\text{\textsc{d}}}$ for $k=1,...,K-1$. Let $b^{th}_{\text{\textsc{d}}}$ a threshold value separating democratic countries from authoritarian ones. Our goal is to rescale the democracy index in the interval $[0,1]$ in order to maintain the confidence bound below the consensus threshold $\varepsilon_c = 0.2$ for democratic countries with $b^k_{\text{\textsc{d}}} >  b^{th}_{\text{\textsc{d}}}$ and above this threshold for authoritarian countries with  $b^k_{\text{\textsc{d}}} <  b^{th}_{\text{\textsc{d}}}$. This can be done by choosing a scale factor $\alpha$ according to the equation:

\begin{equation}
\alpha = \frac{\varepsilon_c}{1- b^{th}_{\text{\textsc{d}}}}
\label{alpha}
\end{equation} 

 In such a way, for simulating the dynamics of a generic country $C_k= \left\{ b^k_{\text{\textsc{pf}}}, b^k_{\text{\textsc{d}}} \right\}$, we will set 

\begin{equation}
\varepsilon^k = (1 - b^k_{\text{\textsc{d}}})\alpha
\label{epsilon}
\end{equation} 

This will ensure that, for a democracy, the higher the value of $b^k_{\text{\textsc{d}}}$, the smaller $\varepsilon^k$ and the more numerous the separate clusters of different opinions produced during the debate process; contrariwise, for more authoritarian regimes, a low value of $b^k_{\text{\textsc{d}}}$ will induce an high value of $\varepsilon^k$ and a collapse towards a single opinion cluster. For sake of clarity, let us describe separately these two different cases.    

\begin{itemize}
\item	
{\it Case $b^k_{\text{\textsc{d}}} > b^{th}_{\text{\textsc{d}}}$: Democratic Countries}. An example of the opinion clustering due to political debate is shown in Figure 1(b) for the same country  $C_k$ presented in panel (a), which has been chosen to be a democratic one. Each cluster has been labelled by its size (i.e. the number of opinions included in it). The configuration reported in this figure is a stationary state for the \hkkp dynamics, since the surviving clusters are made of incompatible opinions, i.e. groups of agents separately synchronized on disjoint opinion profiles, each outside others' compatibility circle. Notice that only a reduced number $N^k_{\text{\textsc{pf}}}=350$ of opinion/agents have been involved in the dynamical process, i.e. have moved from their initial positions in the opinion space - shown in panel (a) -  $\mathbf{x}_i(0), i=1,...,N^k_{\text{\textsc{pf}}}$, to the position of the clusters they belong. We can indicate their shifted opinion profiles as $\mathbf{x}_i(T^k_D), i=1,...,N^k_{\text{\textsc{pf}}}$, being $T^k_D$ the debate time required to obtain such a partial consensus for country $C_k$, measured in terms of the needed parallel simulation steps. It is worth to notice that we consider such a process costless in terms of distortion of citizen's preferences, since we assume that the political debate is a positive feature of free systems through which agents achieve agreements by mutual persuasion. Then, new positions in the opinion space, i.e., clusters, are representative of deliberated compromise.

Once the political debate has been concluded, also in a democratic country the determination of the final decision is required. This requirement is modelled by considering that any institutional framework has its own constitutional mechanisms, which are not being discussed here: we opportunely rescale the confidence bound to the value $\varepsilon^k = 0.5$ and let the \hkkp dynamics start again for all countries with $b^k_{\text{\textsc{d}}} > b^{th}_{\text{\textsc{d}}}$. Thus we do not create artificial differences for different institutional settings. Since $N^k_{\text{\textsc{pf}}}$ clusters represented the non-reducible state of fragmented political agreements, all distances from their positions and the final decision point are considered as the distortion in terms of preferences representation, called 'opinion shift'. For the sake of symmetry, the decision point is approximately depicted as the center of the opinion space. Such a distortion can be evaluated as the length of the path $\mathbf{x}_i(t)$ along which they have to travel from their temporary positions $\mathbf{x}_i(T^k_D)$ to their ultimate position $\mathbf{x}_c(T^k_F)$ (independent of $i$) within the final cluster, being $T^k_F$ the number of time steps needed to get the final decision. Let us define a partial opinion shift for the democratic country $C_k$ as the sum of all these contributions: $O'_k = \sum_{i=1}^{N^k_{\text{\textsc{pf}}}} \mathcal{L}[\mathbf{x}_i(T^k_D), \mathbf{x}_c(T^k_F)]$, being $\mathcal{L}[\mathbf{x}_1, \mathbf{x}_2]$ the length of the (in general non linear) path travelled by a certain agent from $\mathbf{x}_1$ to $\mathbf{x}_2$ and expressed in terms of the unit measure of our opinion space $\mathbf{\Omega}$.

\begin{figure}
	\begin{center}
		\includegraphics[width=4.2in,angle=0]{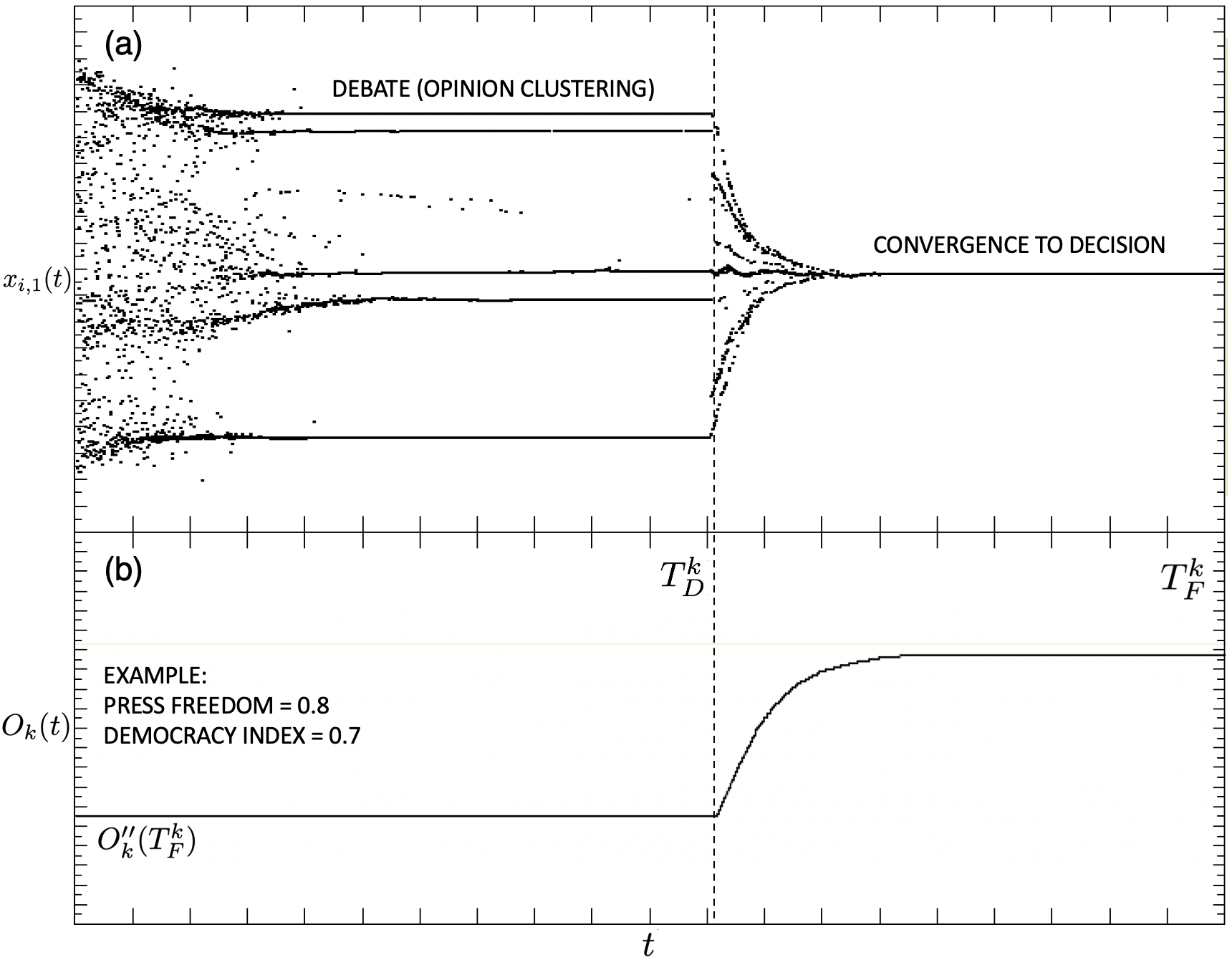}
		\caption{\small (a) An example of time evolution of the $x_{i,1}(t)$ coordinates of the $N^k_{\text{\textsc{pf}}}=400$ opinion profiles for a democratic country with $b^k_{\text{\textsc{pf}}}=0.8$ and $b^k_{\text{\textsc{d}}}=0.7$. A sequential (serial) update of opinions has been assumed for a better visualization. It is clearly visible the transition from the debate phase, with its opinion clustering, to the quick convergence towards the final decision at $T^k_D$. (b) The corresponding time evolution of the global opinion shift $O_k(t)$ shows a sudden increment starting at $T^k_D$, while the final decision is approached.   
		}
		\label{dynamics} 
	\end{center}
\end{figure}

An analogous consideration is valid also for all $N-N^k_{\text{\textsc{pf}}}$ censored agents, remained not included, from the beginning, in the allowed opinion space due to the press freedom restrictions (possibly even worsened in presence of some kind of emergency pressure): since their initial positions $\mathbf{x}_j(0), j=1,...,N-N^k_{\text{\textsc{pf}}}$, remain fixed during the whole dynamical process, we account for such a distortion of preferences by defining another partial opinion shift $O''_k= \sum_{j=1}^{N-N^k_{\text{\textsc{pf}}}} \beta \|\mathbf{x}_j(0)-\mathbf{x}_c(T^k_F)\|$, where each single distance $\|\mathbf{x}_j(0)-\mathbf{x}_c(T^k_F)\|$ is multiplied by a constant factor $\beta>1$, equal for all agents, representing the penalty weight of censorship. Therefore, the final global opinion shift $O_k(T^k_F)$ for the $k$-th democratic country can be written as the sum of the two contributions $O'_k+O''_k$: 

\begin{equation}
O_k (T^k_F) = \sum_{i=1}^{N^k_{\text{\textsc{pf}}}} \mathcal{L}[\mathbf{x}_i(T^k_D), \mathbf{x}_c(T^k_F)] + \beta \sum_{j=1}^{N-N^k_{\text{\textsc{pf}}}} \|\mathbf{x}_j(0)-\mathbf{x}_c(T^k_F)\| 
\label{o_sum}
\end{equation} 

Our results will be presented in terms of the {\it per capita} global opinion shift: $o_k = O_k / N$, which is the first output key variable of the model and will be also called  \textsc{gos}.        

The final, stationary, configuration is reported in Figure 1(c) for the same democratic country that we are considering in this example: it clearly appears that the $N^k_{\text{\textsc{pf}}}=350$ opinion points have now collapsed in the central cluster. The total decision time $T^k_F$ needed to reach such a final configuration for the considered country will be our second output key variable. It will be also called  \textsc{tdt}. Differences in the total decision time, according to the situation in the country at hand, explain why the institutional setting of different countries can manifest different efficiency in policy action. 

In Figure 2 an example of the whole dynamical process is depicted for a generic country with $b^k_{\text{\textsc{pf}}}=0.8$ and $b^k_{\text{\textsc{d}}}=0.7$. In panel (a), the time evolution of the $x_{i,1}(t)$ coordinates of the opinion profiles, $i=1,...,N^k_{\text{\textsc{pf}}}$, shows the initial formation of the clusters due to the political debate which, after $t=T^k_D$ simulation steps, start to converge towards the final decision taken at $t=T^k_F$. Correspondently, as shown in panel (b), the global opinion shift $O_k(t)$, starting from the reference value $O''_k(T^k_F)$, increases for $t>T^k_D$ due to the further distortion of preferences represented by the term $O'_k(t)$. 

\item 
{\it Case $b^k_{\text{\textsc{d}}} < b^{th}_{\text{\textsc{d}}}$: Authoritarian Countries}.
For these countries the \hkkp dynamics is simplified by the fact that, as previously said, they do not allow a debate process inside themselves. This is automatically obtained by the requirement $b^k_{\text{\textsc{d}}} < b^{th}_{\text{\textsc{d}}}$ which implies, for a given authoritarian country $C_k= \left\{ b^k_{\text{\textsc{pf}}}, b^k_{\text{\textsc{d}}} \right\}$, a confidence bound $\varepsilon^k$ above the consensus threshold (see Eq.\ref{epsilon}). Therefore, starting from the usual initial conditions at $t=0$ $\mathbf{x}_i(0), i=1,...,N^k_{\text{\textsc{pf}}}$, for $t>0$, all the opinions included in the allowed space, reduced by the restrictions on press freedom, will now directly collapse towards a single cluster placed at the central position $\mathbf{x}_c(T^k)$ representing the final decision. $T^k$ is here the total decision time,  \textsc{tdt}, needed for the dynamics to reach such a stationary state, which essentially is analogous to that one shown in Figure 1(c) for the democratic country. The corresponding behavior of the $x_{i,1}(t)$ in the opinion space will also be similar to that shown in Figure 2(a) for $t>T^k_D$.   

Also in this case, like in the democratic one, is it possible to define the global opinion shift as the sum of two terms: $O_k = \sum_i^{N^k_{\text{\textsc{pf}}}} \mathcal{L}[\mathbf{x}_i(0), \mathbf{x}_c(T^k_F)] + \beta \sum_j^{N-N^k_{\text{\textsc{pf}}}} \|\mathbf{x}_j(0)-\mathbf{x}_c(T^k_F)\|$, concerning the contribution of both the $N^k_{\text{\textsc{pf}}}$ agents involved in the dynamical process and the $N-N^k_{\text{\textsc{pf}}}$ agents censored by the political regime, which of course are expected to be here more numerous than for the democratic countries. The penalty factor $\beta$ hase the same value than in the previous case. Again, $o_k = O_k / N$ will be the corresponding {\it per capita} global opinion shift,  \textsc{gos}, which will be adopted for presenting the simulation results.

\end{itemize}

\section{Case study and simulations results}

In the previous section we introduced the \hkkp model, describing in detail its dynamical aspects with reference to a generic set of ideal countries $C_k= \left\{ b^k_{\text{\textsc{pf}}}, b^k_{\text{\textsc{d}}} \right\}$. Let us now present a sample of $K$ real countries, considered as a case-study. For each country we identify two real indexes able to give back a classification of the countries in terms of their two main parameters, the press freedom index $b^k_{\text{\textsc{pf}}}$ and the democracy index $b^k_{\text{\textsc{d}}}$. Our scope will be to statistically evaluate, with the help of extended numerical simulations, the performance of each country in terms of both  \textsc{gos} and  \textsc{tdt}, and to highlight how the emergency pressure may asymmetrically affect the efficiency of the decision process in different contexts.

\subsection{Sample selection}

Political regime is the key characteristic to set institutional framework. Although there is not a universally accepted criteria, there seems to be a general agreement on a clear distinction between democracy and dictatorship as polar types of systems. Based on such a distinction, Cheibub et al. (2010) identify a six-fold classification of political regimes: parliamentary democracy; semi-presidential democracy; presidential democracy; monarchic dictatorship; civil dictatorship; and military dictatorship. 
According with this classification, we select a sample of $K=20$ countries $C_k$ ($k=1,...,K$) to work with, each of them defined by the following two real indexes. 

The first index, which is assumed as press freedom parameter $b^k_{\text{\textsc{pf}}}$, is the World Press Freedom Index as published yearly since 2002 by Journalists Sans Frontiers. It ranks 180 countries and regions according to the level of freedom available to journalists and it is based on an evaluation of pluralism, independence of the media, quality of legislative framework and safety of journalists in each country and region.  

The second index, assumed as democracy parameter $b^k_{\text{\textsc{D}}}$, is the Democracy Index calculated by Economist Intelligence Unit, a UK-based company. It is a composite index that integrates four distinct components. The first one aims at representing the political institutions focusing on different aspects: competitive, multi-party political system; universal adult suffrage; regularly contested elections conducted based on secret ballots, reasonable ballot security and the absence of massive voter fraud; and significant public access of major political parties to the electorate through the media and generally open political campaigning. The complementary three other factors aim at integrating the scale from democracies to dictatorships. More specifically, they consist of indicators that are assumed to shorten the distance between formal rules and the actual working of systems. The Economist democracy index allows to collapse the Cheibub's six-fold classification into four broader groups: full democracies (FUD), flawed democracies (FLD), hybrid regimes (HYB) and authoritarian (AUT).  

\begin{table} 
	\centering%
	\begin{tabular}[c]{l|l|l|l|l|r}
		\textsc{regime} &\textsc{EC}&\textsc{country} & $b^k_{\text{\textsc{pf}}}$&$b^k_{\text{\textsc{d}}}$ &$\bar{N}^k_{\text{\textsc{pf}}}$ \\ \hline \hline
		Parliamentary democracy 
		& FUD &\textsc{germany}          & $0,878$ & $0,868$  &  439 \\
		& FLD &\textsc{india}                   & $0,547$  & $0,690$ &  273 \\
		& FLD &\textsc{japan}                 & $0,711$   & $0,799$ &  356 \\
		& FLD & \textsc{italy}                    &$0,763$   &$0,752$ &  382 \\
		& FUD &\textsc{sweden}            &$0,908$   &$0,939$ &  454 \\
		& FUD &\textsc{uk}                        &$0,771$    & $0,852$ &  385 \\
		Semi-presidential democracy
		& FUD &\textsc{canada}             & $0,847$ & $0,922 $  &  424 \\
		& FUD &\textsc{finland}              & $0,921$  & $0,925$ &  460 \\
		& FUD &\textsc{france}               & $0,771$  & $0,812$    &  385 \\
		Presidential democracy
		& FLD &\textsc{argentina}         &$0,712$   & $0,702$   &  356 \\
		& FLD &\textsc{brazil}                  & $0,660$  & $0,686$  &  330 \\
		& FLD &\textsc{south korea}     & $0,763$  &$0,800$ &  382 \\
		& FLD &\textsc{usa}                     & $0,762$  & $0,796$  &  381 \\
		& HYB &\textsc{turkey}               & $0,500$  & $0,409$  &  350 \\
		Civil dictatorship
		& AUT &\textsc{china}                 &$0,215$  & $0,226$   &  108 \\
		& AUT &\textsc{kazakistan}       & $0,459$	& $0,294$ &  229 \\
		& AUT &\textsc{russia}                & $0,511$    & $0,311$  &  255 \\
		Military dictatorship
		& AUT &\textsc{iran}                     & $0,352$ & $0,238$ &  176 \\
		Monarchic dictatorship
		& AUT &\textsc{saudi arabia}   & $0,379$  & $0,193$  &  189 \\
		& AUT &\textsc{un.arab em.}     &$0,573$   &$0,276$ &  287 \\
		\hline \hline
	\end{tabular}
	\caption{Parameters setting for the 20 countries: the Cheibub's regime classifications (column I); the classification by the Economist (column II, where full democracies are indicated by FUD, flawed democracies by FLD, hybrid regimes by HYB and authoritarian ones by AUT); the names of countries (column III); the press freedom index $b^k_{\text{\textsc{pf}}}$(column IV); the democracy index $b^k_{\text{\textsc{d}}}$ (column V); the asymptotic reduced number of opinions participating to the dynamics, $\bar{N}^k_{\text{\textsc{pf}}} \sim N \cdot b^k_{\text{\textsc{pf}}}$ for $N=500$ (column VI).}%
	\label{tabella-parametri}%
\end{table}

\begin{figure}
	\begin{center}
		\includegraphics[width=4.7in,angle=0]{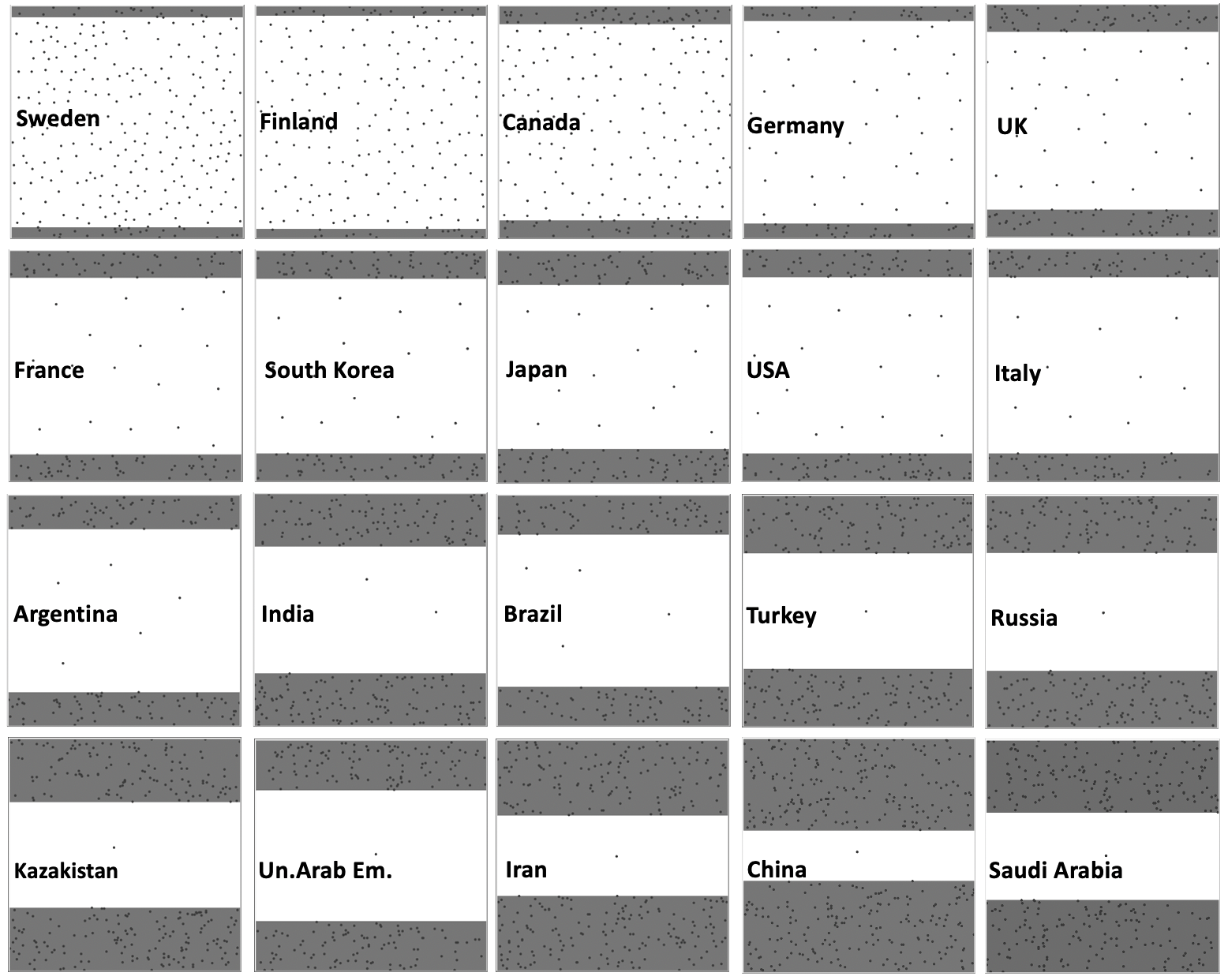}
		\caption{\small The opinion spaces of the $20$ selected countries as they appear at time $T^k_D$, for democratic countries, and at time $T^k_F$ for authoritarian countries. The different degrees of restriction of the allowed space (in white) due to a limited press freedom are visible.   
		}
		\label{countries} 
	\end{center}
\end{figure}

Table \ref{tabella-parametri} summarizes sample composition and parameters setting for the selected countries, organized according to the Cheibub classification. Acronymous of the Economist classification (EC) are also reported for comparison. Values of $b^k_{\text{\textsc{pf}}}$ and $b^k_{\text{\textsc{D}}}$ have been normalized in the interval $[0,1]$. Notice that, as one could expect, real countries show a certain correlation between the two indexes: strong democracies (such as Sweden or Finland) tend to have quite high values for both of them, while the most authoritarian regimes (like China or Saudi Arabia) show a couple of lower values. Anyway, between these extremes, a variegated spectrum of cases can be found, thus it will be interesting to see how the (non linear) interplay between these two indexes will impact on the outcomes of the simulations. As discussed before, only a reduced number of opinion/agents will be involved in the dynamic process, depending on the freedom index. Hence, for each country, we also report the asymptotic number of opinions participating to the dynamics, $\bar{N}_{PF}^k \sim Nb_{PF}^k$ for $N=500$. The corresponding restricted opinion space allowed for the dynamics is shown in Figure \ref{countries}: for each country, censored agents - in the grey area - do not participate to the dynamics, whereas the involved ones - in the white area - converge either towards many clusters (in democracies) or to just a single cluster (in authoritarian regimes).

In Table \ref{debate-index-list}, following the procedure explained in section 2.2, we report the same $20$ countries ordered by decreasing values of $b_D^k$ together with the corresponding value of the confidence bound $\varepsilon^k$ calculated according to equations (4) and (5), where the threshold value $b_D^{th}=0.6$ separating flawed democracies from hybrid regimes (highlighted in grey) directly derives from the ranking proposed by the Economist.

\begin{table} 
	\centering
	\begin{tabular}[c]{l|c|c|}
		\textsc{country}&$b^k_{\text{\textsc{d}}}$ & $\varepsilon^k$ \\ \hline \hline
		\textsc{sweden}  &$0,939$&  $0,031$  \\
		\textsc{finland}  &$0,925$ &    $0,038$ \\
		\textsc{canada}  &$0,922$ &   $0,039$  \\
		\textsc{germany}  &$0,868$&   $0,066$   \\
		\textsc{uk}  &$0,852$&    $0,074$  \\
		\textsc{france}  &$0,812$&    $0,094$  \\
		\textsc{south korea}  &$0,800$ &  $0,100$  \\
		\textsc{japan}  &$0,799$&   $0,101$   \\
		\textsc{usa}  &$0,796$  &   $0,102$ \\
		\textsc{italy}  &$0,752$  &  $0,124$ \\
		\hline \hline
	\end{tabular}
	\begin{tabular}[c]{|l|c|c}
		\textsc{country}&$b^k_{\text{\textsc{d}}}$ & $\varepsilon^k$ \\ \hline \hline
		\textsc{argentina}  &$0,702$ &    $0,149$ \\
		\textsc{india}  &$0,690$&    $0,155$  \\
		\textsc{brazil}  &$0,686$ &  $0,157$ \\
		\rowcolor[gray]{0.9}
		\textsc{turkey}  &$0,409$  &   $0,296$ \\
		\rowcolor[gray]{0.9}
		\textsc{russia}  &$0,311$ &    $0,345$ \\
		\rowcolor[gray]{0.9}
		\textsc{kazakistan}  &$0,294$  &  $0,353$ \\
		\rowcolor[gray]{0.9}
		\textsc{un.arab em.}  &$0,276$ &   $0,362$ \\
		\rowcolor[gray]{0.9}
		\textsc{iran}  &$0,238$  &    $0,381$\\
		\rowcolor[gray]{0.9}
		\textsc{china}  &$0,226$ &    $0,387$ \\
		\rowcolor[gray]{0.9}
		\textsc{saudi arabia}  &$0,193$  &   $0,404$ \\
		\hline \hline
	\end{tabular}
	\caption{Countries ordered by decreasing values of the democracy index $b_D^k$. For each country, the corresponding value of the confidence bound $\varepsilon^k$ is also reported. The threshold value $b_D^{th}=0.6$ (corresponding to the critical value $\varepsilon_c=0.2$ of the confidence bound) clearly separates full and flawed democracies from hybrid and authoritarian regimes (highlighted in grey), according with the Economist classification.
	}
	\label{debate-index-list}
\end{table}

\subsection{Results of simulations}

Two different scenarios for the presented countries will be now analyzed through our agent-based simulations: the benchmark scenario without emergency $(e=0)$ and an emergency scenario in which different emergency pressure levels will be tested $(e>0)$.

\subsubsection{Emergency pressure $e=0$}

\begin{figure}
	\begin{center}
		\includegraphics[width=5in,angle=0]{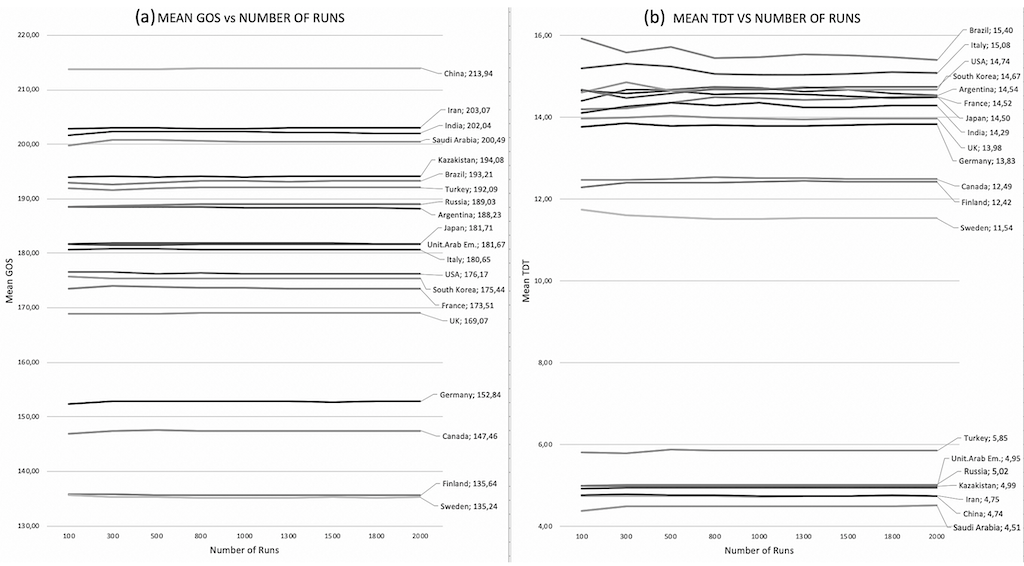}
		\caption{\small Stability test for both Mean  \textsc{gos} (a) and Mean  \textsc{tdt} (b) as function of the number of simulation runs over which the averages have been performed. It clearly appears that both the considered quantities are stable enough already after $1000$ runs.}
		\label{stability} 
	\end{center}
\end{figure}

The benchmark scenario corresponds to no emergency pressure, so that it results only dependent on countries' parameters. As illustrated before, the model focuses on two key outcomes:  \textsc{gos}, as measured by the {\it per capita} distance between the original opinions of agents and their final decision, and the  \textsc{tdt}, i.e. the time necessary for the country to converge towards a final decision. First, we need to perform a stability test for both mean  \textsc{gos} (a) and mean  \textsc{tdt} (b) as a function of the number of simulation runs over which the averages are computed. Both considered quantities appear to be stable enough already after 1,000 runs, as shown in Figure \ref{stability}.

Table \ref{mean-gos-tdt} shows the average values of both  \textsc{gos} and  \textsc{tdt} over 1000 runs. From a comparison with Table \ref{debate-index-list} we immediately notice that the  \textsc{gos} ranking is able to well reproduce the ranking of the real democracy index $b_D^k$, thus confirming the goodness of the hypothesis at the basis of the \hkkp model. Moreover, as expected, on average lower values of  \textsc{gos} are associated to higher values of  \textsc{tdt}. This is, in general, the case of real democratic countries, where the political debate guarantees final decisions closer to opinions, but requiring more time to reach them. The opposite occurs for dictatorships, whose values of  \textsc{gos} are higher but  \textsc{tdt} is much lower: in fact, real authoritarian regimes are typically quick to decide, but opinions have no status. However, it is interesting to notice that, immediately after the jump in  \textsc{tdt} from dictatorships to democracies (from 5.6 to 11.52), we found a small group of strong democracies, with a very small  \textsc{gos} but also with a relatively low  \textsc{tdt}, compared with the other democracies: this is a typical non linear outcome of the \hkkp model, which favors a quick aggregation of small clusters in the debate phase of countries with a very high democracy index (i.e. a small confidence bound) while hinders a rapid convergence in big clusters for democracies with a $b_D^k \gtrapprox b_D^{th}$ (i.e. with a confidence bound just below the critical threshold $\epsilon_c$), thus enhancing their  \textsc{tdt}. We think that this emergent properties of our model well captures the real behavior of flawed democracies, whose bureaucratic apparatus slows down the political decision-making processes.  

\begin{table} 
	\centering%
	\begin{tabular}
		[c]{l|l|}%
		\textsc{country} &\textsc{gos} \\ 
		\hline \hline
		{\small 	\textsc{sweden}    }     & {\small $135,15$ }\\
		{\small 	\textsc{finland}        }     & {\small $135,65$ } \\
		{\small    \textsc{canada}       }     & {\small $147,43$ } \\
		{\small   \textsc{germany}   }     & {\small $152,85$ } \\
		{\small   \textsc{uk}                 }     & {\small $169,03$ } \\
		{\small \textsc{france}         }     & {\small $173,66$ } \\
		{\small   \textsc{south korea}}   & {\small $175,40$ } \\
		{\small   \textsc{usa}                }    & {\small $176,25$  }\\
		{\small   \textsc{italy}               }    & {\small $180,75$}  \\
		{\small  \textsc{un.arab em.}  } & {\small $181,67$}  \\
		{\small   \textsc{japan}             }    & {\small $181,86$ } \\
		{\small   \textsc{argentina}      }   & {\small $188,41$}  \\
		{\small  \textsc{russia}            }    & {\small $189,01$ } \\
		{\small  \textsc{turkey}            }   & {\small $192,12$ } \\
		{\small  \textsc{brazil}              }  & {\small $193,30$  }\\
		{\small  \textsc{kazakistan}    }   & {\small $194,01$}  \\
		{\small   \textsc{saudi arabia} }   & {\small $200,53$ } \\
		{\small   \textsc{india}               }    & {\small $202,34$} \\
		{\small  \textsc{iran}                 }    & {\small $203,90$ } \\
		{\small  \textsc{china}              }   & {\small $213,90$ } \\
		\hline \hline
	\end{tabular}
	\begin{tabular}
		[c]{|l|l|}%
		\textsc{country} &\textsc{tdt} \\ 
		\hline \hline
		{\small 	\textsc{saudi arabia}     }     &{\small \, $4,50$ } \\
		{\small    \textsc{china}              }          &{\small \, $4,74$ } \\
		{\small    \textsc{iran}                 }          &{\small \, $4,75$  }\\
		{\small     \textsc{un.arab em.}    }      &{\small \, $4,95$ } \\
		{\small     \textsc{kazakistan}        }     &{\small \, $4,99$  }\\
		{\small    \textsc{russia}          }            &{\small \, $5,02$ } \\
		{\small      \textsc{turkey}          }           &{\small \, $5,86$ } \\
		{\small     \textsc{sweden}      }            &{\small  $11,52$ } \\
		{\small      \textsc{finland}         }            &{\small  $12,43$ } \\
		{\small    \textsc{canada}         }           &{\small  $12,53$}  \\
		{\small      \textsc{germany}      }          & {\small $13,79$ } \\
		{\small        \textsc{uk}                  }            &{\small  $13,97$ } \\
		{\small    \textsc{india}             }            &{\small  $14,35$ } \\
		{\small    \textsc{france}             }          & {\small $14,46$ } \\
		{\small     \textsc{japan}            }         &{\small $14,58$  }\\
		{\small     \textsc{usa}        }                  & {\small $14,68$}  \\
		{\small      \textsc{south korea}   }      & {\small $14,69$}  \\
		{\small    \textsc{argentina}     }       & {\small $14,72$ } \\
		{\small    \textsc{italy}                }         & {\small $15,05$ } \\
		{\small     \textsc{brazil}             }         & {\small $15,47$ } \\
		\hline \hline
	\end{tabular}
	\begin{tabular}
		[c]{|l|l|l|l}%
		\textsc{country} &\textsc{g/t}& \textsc{gos}& \textsc{tdt}  \\ 
		\hline \hline
		{\small 		\textsc{finland}      }    & {\small $10,91$}& {\small $135,65$ }&{\small $12,43$ }   \\
		{\small     \textsc{germany}    }      & {\small $11,08$}& {\small $152,85$} &{\small $13,79$ }   \\
		{\small     \textsc{sweden}    }      & {\small $11,73$}& {\small $135,15$} &{\small $11,52$  }  \\
		{\small     \textsc{canada}    }      & {\small $11,77$}&{\small  $147,43$}&{\small $12,53$  }  \\
		{\small     \textsc{south korea}     }     & {\small $11,94$}& {\small $175,40$} &{\small $14,69$  }  \\
		{\small     \textsc{usa}      }    & {\small $12,01$}& {\small $176,25$} &{\small $14,68$ }   \\
		{\small     \textsc{france}   }       & {\small $12,01$}&{\small  $173,66$ }&{\small $14,46$}    \\
		{\small     \textsc{italy}     }     & {\small $12,01$}&{\small  $180,75$ }&{\small $15,05$ }   \\
		{\small     \textsc{uk}       }   & {\small $12,10$}& {\small $169,03$} &{\small $13,97$ }   \\
		{\small     \textsc{japan}    }      & {\small $12,47$}& {\small $181,86$} &{\small $14,58$}    \\
		{\small     \textsc{brazil}    }      & {\small $12,50$}& {\small $193,30$} &{\small $15,47$ }   \\
		{\small     \textsc{argentina}      }    &{\small  $12,80$}& {\small $188,41$} &{\small $14,72$ }   \\
		{\small     \textsc{india}       }   &{\small  $14,10$}& {\small $202,34$ }&{\small $14,35$ }   \\
		{\small     \textsc{turkey}      }    &{\small  $32,77$}& {\small $192,12$ }&{\small \, $5,86$  }  \\
		{\small    \textsc{un.arab em.}    }      & {\small $36,71$}& {\small $181,67$} &{\small \, $4,95$  }  \\
		{\small     \textsc{russia}    }      & {\small $37,63$}& {\small $189,01$ }&{\small \, $5,02$}    \\
		{\small    \textsc{kazakistan}    }      &{\small  $38,86$}& {\small $194,01$} &{\small \, $4,99$ }   \\
		{\small    \textsc{iran}      }    & {\small $42,71$}& {\small $202,90$ }&{\small \, $4,75$ }   \\
		{\small   \textsc{saudi arabia}       }   &{\small  $44,54$}& {\small $200,53$ }&{\small \, $4,50$  }  \\
		{\small    \textsc{china}       }   & {\small $45,14$}&{\small $213,90$} &{\small \, $4,74$}    \\
		\hline \hline
	\end{tabular}
	\caption{\small Rankings of averages over 1000 runs. \textsc{gos} (left panel), \textsc{tdt} (mid panel), and \textsc{g/t} ratio (right panel - \textsc{gos} and \textsc{tdt} are repeated for reading convenience).}
	\label{mean-gos-tdt}%
\end{table}

\begin{figure}
	\begin{center}
		\includegraphics[width=4.5in,angle=0]{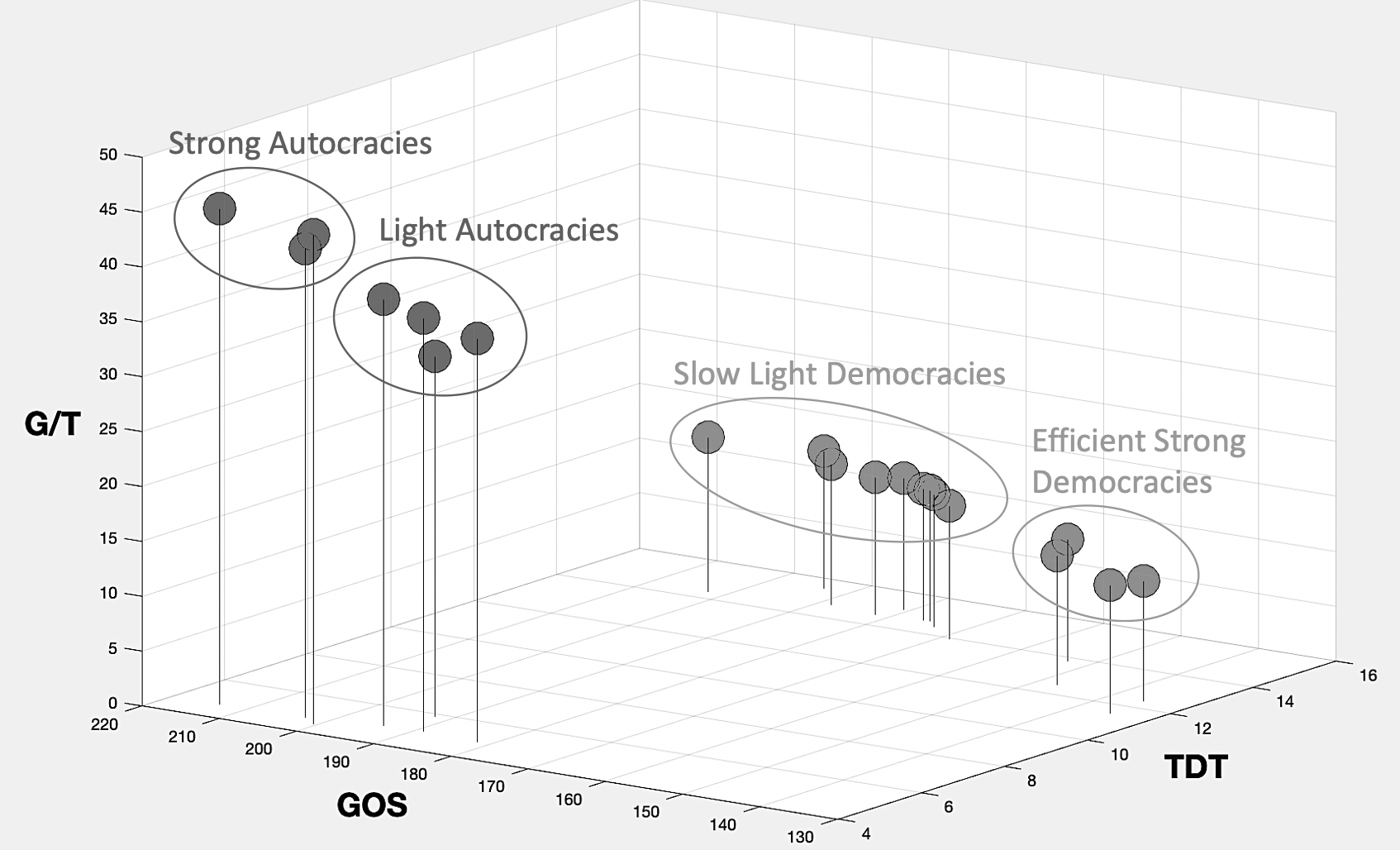}
		\caption{\small A 3D plot where each of the considered $20$ countries is reported like a point as function of the three main indicators of performance of the \hkkp model,  \textsc{gos},  \textsc{tdt} and their ratio  \textsc{g/t}, averaged over $1000$ runs. Through a comparison with Table 3 it is easy to identify each single country (see text for more details). Four clusters of countries can be clearly distinguished, which we have labelled as: strong and light autocracies (characterized by a small \textsc{tdt} but high values for both  \textsc{gos} and  \textsc{g/t}); slow light democracies (with quite high  \textsc{gos} and  \textsc{tdt}, but low  \textsc{g/t}); efficient strong democracies (with low \textsc{gos} and \textsc{g/t} and also a lower  \textsc{tdt} than other democracies).}
		\label{3d_fig} 
	\end{center}
\end{figure}

It is therefore evident the necessity to combine both  \textsc{gos} and  \textsc{tdt} information into a unique synthetic index, in order to build a global ranking which could take into account all the above considerations. To this aim we calculate the  \textsc{gos}/\textsc{tdt} ratio (\textsc{g/t}), i.e. the cost in terms of distance between original opinion and final decision for each time decision unit. In other words, how costly is to be quick. Values range between $10.91$ to $14.10$ for democracies, while dictatorships range from $32.77$ to $45.14$. The introduction of this index, combined with previous information, allows to identify the existence of four different groups of countries: efficient strong democracies (Finland, Germany, Sweden, Canada) with low  \textsc{gos} and  \textsc{tdt} (compared to other democracies) and with low  \textsc{g/t}; slow flawed democracies (the remaining democratic countries, from South Korea to India), with quite high \textsc{gos} and  \textsc{tdt}, but low  \textsc{g/t}; strong dictatorships (Iran, Sudi Arabia, China), characterized by small  \textsc{tdt} and the highest values of  \textsc{gos} and  \textsc{g/t}; light dictatorships (Kazakistan, Russia, United Arab Emirates and Turkey), with a  \textsc{tdt} comparable with that of strong dictatorships but with a relatively smaller  \textsc{gos} and  \textsc{g/t}. The situation is summarized in Figure \ref{3d_fig}, which shows a $3$D plot where each one of the considered $20$ countries is reported like a point as function of the three main indicators of performance of the \hkkp model,  \textsc{gos},  \textsc{tdt} and their ratio  \textsc{g/t}, averaged over $1000$ runs. We obtain four groups, highlighted with ellipsoidal curves for better identification, which we have labelled as strong and light autocracies (characterized by a small \textsc{tdt} but high values for both  \textsc{gos} and  \textsc{g/t}); slow light democracies (with quite high  \textsc{gos} and  \textsc{tdt}, but low  \textsc{g/t}); efficient strong democracies (with low \textsc{gos} and \textsc{g/t} and also a lower  \textsc{tdt} than other democracies). Notice that the distribution of countries in such groups is in a good agreement with the classification of Economist presented in Table \ref{tabella-parametri}. 

\subsubsection{Emergency pressure $e>0$}

Taking into account the two parameters of decision, i.e. the press freedom and the democracy index, in our simulation clearly emerges that being democratic is time-costly since opinions need time to converge towards a common decision. Dictatorships are much quicker to take a decision, however, the lack of opinion representativeness causes an increase of  \textsc{gos} which could over-compensate the benefits of speed. But what does it happen when systems in which decisions occur are stressed by some kind of emergency pressure? In order to test such a condition we introduce different levels of emergency pressure, ranging from $10\%$ to $80\%$. As previously explained, the way in which emergency pressure works in our model is by adding a further restriction to the allowed opinion space, which enhances the weight of censored agents in the global opinion shift.

\begin{figure}
	\begin{center}
		\includegraphics[width=5in,angle=0]{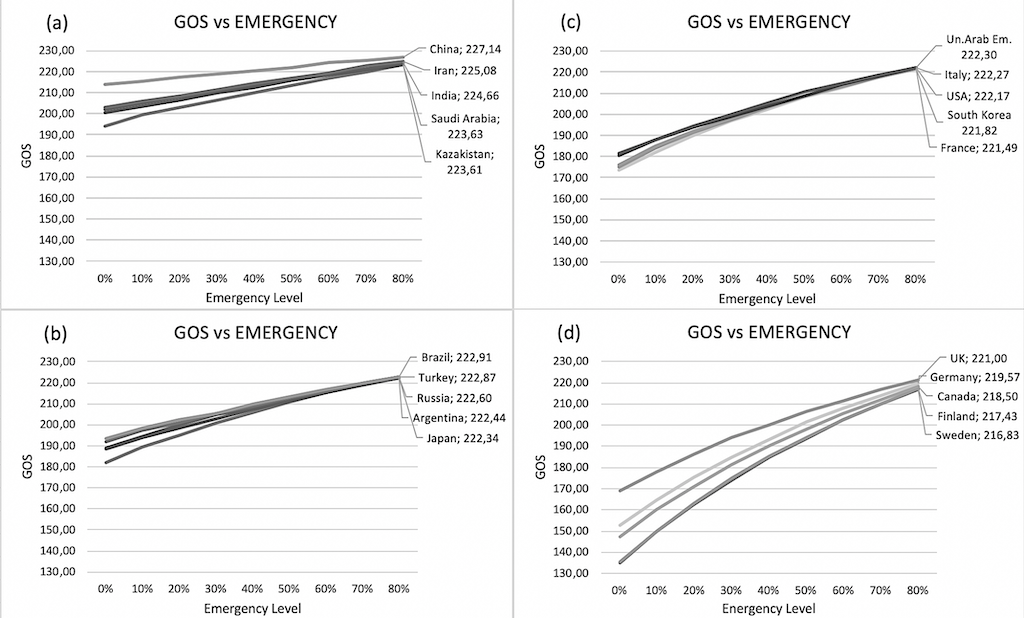}
		\caption{\small Emergency response of the $20$ countries in terms of  \textsc{gos} as function of an increasing level of alarm. Countries are organized in four groups, from (a) to (d), ordered according to decreasing  \textsc{gos} values. Averages are over $1000$ runs.}
		\label{emergencyGOS} 
	\end{center}
\end{figure}

Figure \ref{emergencyGOS} shows the emergency responses of the $20$ countries in terms of  \textsc{gos} as function of an increasing level of emergency pressure. Countries are organized in four groups, from (a) to (d), according to decreasing  \textsc{gos} values. 

\begin{figure}
	\begin{center}
		\includegraphics[width=5in,angle=0]{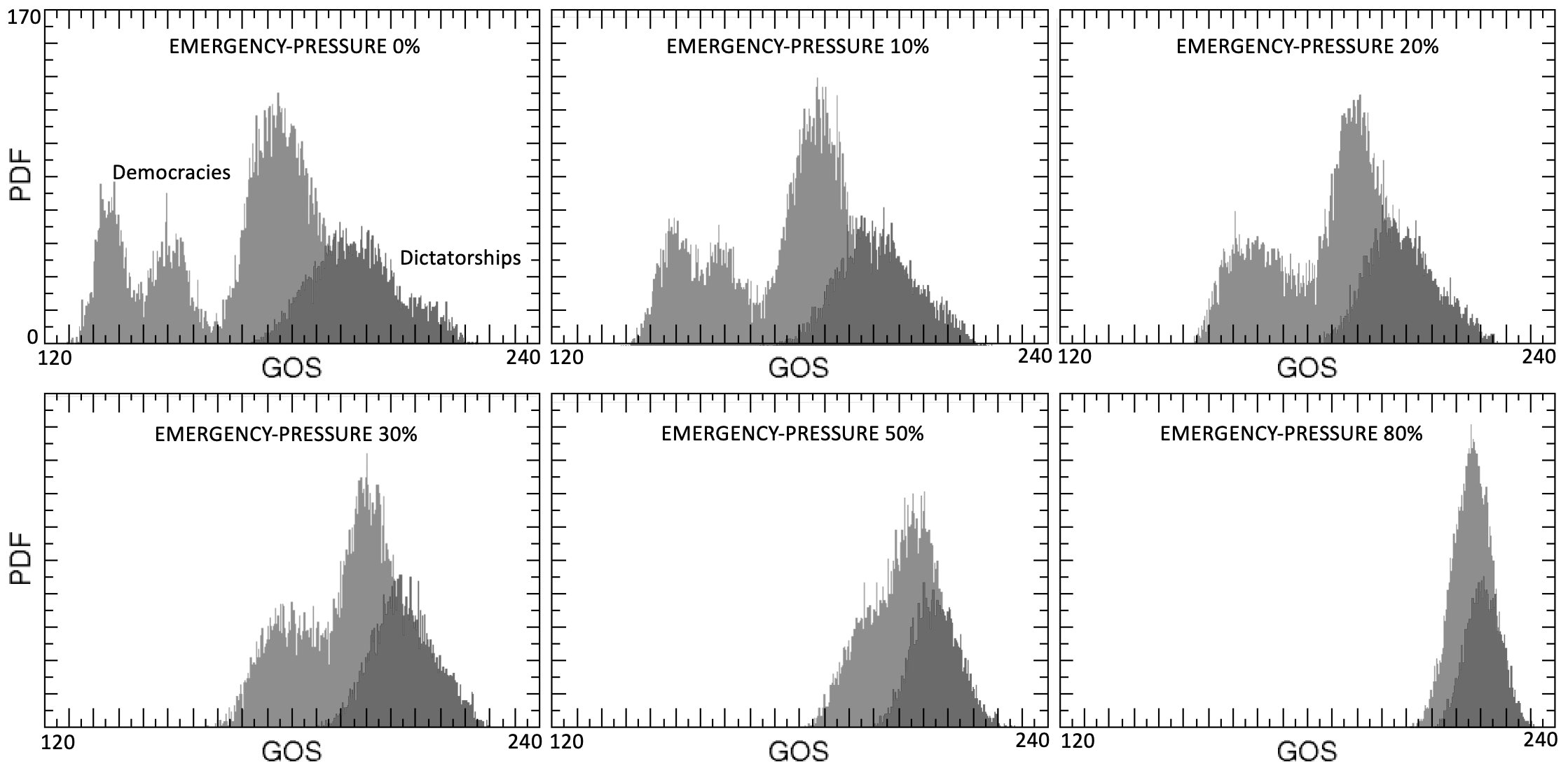}
		\caption{\small Behaviour of the  \textsc{gos} distributions, cumulated for democratic countries and for authoritarian ones, against an increasing level of emergency pressure. It is evident the progressive shift of democracies towards the high degree of distortion of preferences typical of authoritarian regimes under the effect of the increasing level of emergency.}
		\label{GOS-distributions} 
	\end{center}
\end{figure}

Regardless of regimes, moving from low towards high emergency pressure levels all countries in each group show convergent trajectories toward a common point with a higher  \textsc{gos} value. This means that, under the effect of increasing emergency levels, there exists a progressive shift of democracies towards the higher degree of distortions of preferences typical of dictatorship regimes. A visual inspection of Figure \ref{GOS-distributions}, which illustrates the behavior of the  \textsc{gos} distributions cumulated for democracy countries and for dictatorial ones, confirms the dynamic of this convergent shift as a response to the increasing emergency pressure.

\begin{figure}
	\begin{center}
		\includegraphics[width=5in,angle=0]{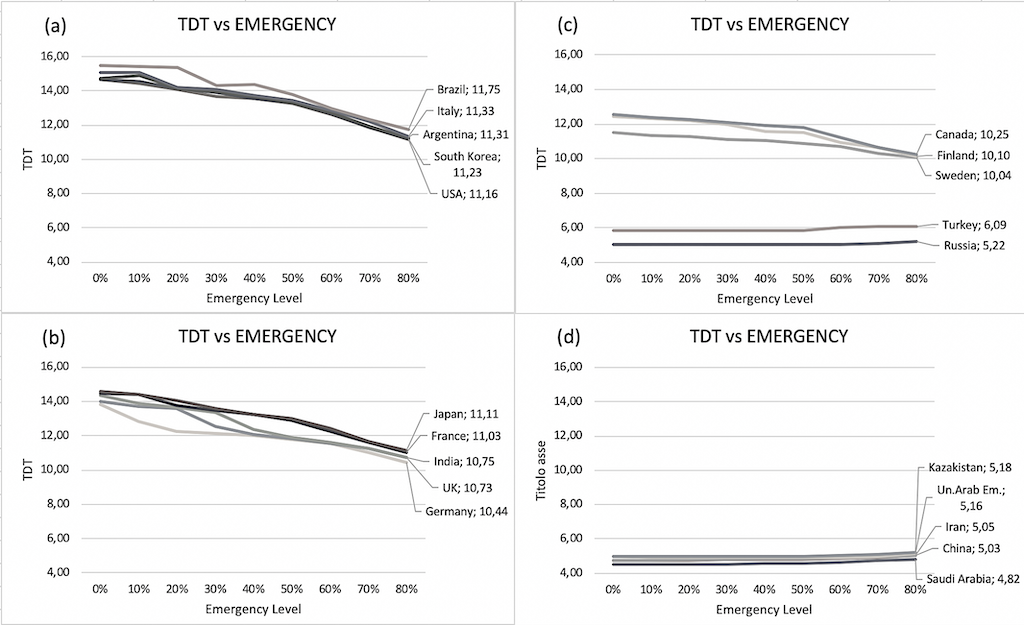}
		\caption{\small Emergency response of the $20$ countries in terms of  \textsc{tdt} as function of an increasing level of alarm. Countries are organized in four groups, from (a) to (d), ordered according to decreasing  \textsc{tdt} values. Averages are over $1000$ runs.}
		\label{emergencyTDT} 
	\end{center}
\end{figure}

The second outcome to look at is the emergency response in terms of  \textsc{tdt}, as shown in Figure \ref{emergencyTDT}. It is clearly visible that emergency pressure forces  \textsc{tdt} to reduce in the case of democracies, while it seems to be irrelevant in the case of dictatorships, which evidently already operated as in emergency conditions. However, although speed to converge does not change for these authoritarian regimes, as previously observed emergency still increase the distance between opinions and final decisions. 

\begin{figure}
	\begin{center}
		\includegraphics[width=5in,angle=0]{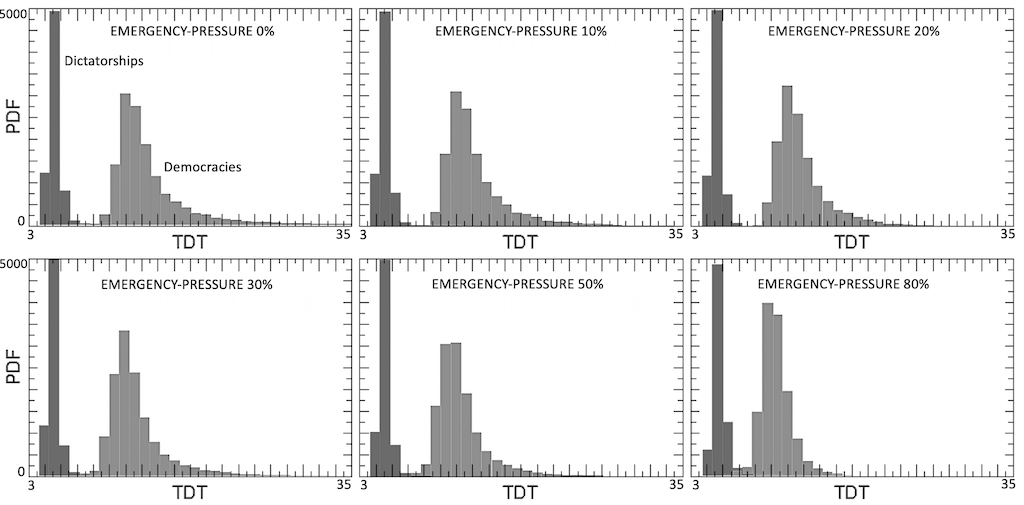}
		\caption{\small Behavior of the  \textsc{tdt} distributions, cumulated for democratic countries and for authoritarian ones, against an increasing level of emergency pressure. Again, democracies tend to approach the short decision time windows characteristic of authoritarian regimes as the emergency increases.}
		\label{TDT-distribution} 
	\end{center}
\end{figure}

Analogously to what illustrated before, Figure \ref{TDT-distribution} shows the behavior of the  \textsc{tdt} distributions, cumulated for democracy countries and for dictatorial ones, under an increasing emergency pressure. In this case, a small progressive tendency of democracies to reduce time to decide can be appreciated, but the distance from dictatorships remains persistent even at the maximum emergency level.

Moving to the synthesis index, the average  \textsc{g/t} ratio confirms the movement of democracies under emergency pressure toward dictatorships. Ranging from $10$ percent to $80$ percent of emergency pressure, Table \ref{g/t-ranking-emergency} displays the  \textsc{g/t} ratio for all countries in the sample. Despite the ranking does not change, differences among countries are affected and democracies converge to position close to dictatorship.

\begin{table} 
	\centering%
	\begin{tabular}
		[c]{l|c|c|c|c|c|c}%
		\textsc{country} & $e=0\%$ & $e=10\%$ & $e=20\%$ & $e=30\%$ & $e=50\%$ &  $e=80\%$   \\ 
		\hline \hline 
		\textsc{finland}               & $10,91$ & $12,00$ & $12,891$ & $13,83$ & $15,17$ & $18,69$  \\
		\textsc{germany}           & $11,08$ & $12,17$ & $13,01$ & $13,96$ & $15,49$ & $19,50$  \\
		\textsc{sweden}             & $11,73$ & $12,49$ & $13,40$ & $14,04$ & $15,52$ & $19,59$  \\
		\textsc{canada}              & $11,77$ & $12,59$ & $13,46$ & $14,17$ & $15,65$ & $19,80$  \\
		\textsc{south korea}& $11,94$ & $12,70$ & $13,47$ & $14,19$ & $15,80$ & $19,92$  \\
		\textsc{usa}                     & $12,01$ & $12,79$ & $13,57$ & $14,41$ & $15,88$ & $20,02$  \\
		\textsc{france}               & $12,01$ & $12,84$ & $13,65$ & $14,47$ & $16,03$ & $20,15$  \\
		\textsc{italy}                    & $12,01$ & $12,93$ & $13,80$ & $14,67$ & $16,47$ & $20,26$  \\
		\textsc{uk}                       & $12,10$ & $12,93$ & $13,81$ & $14,95$ & $16,80$ & $20,33$  \\
		\textsc{japan}                 & $12,47$ & $13,02$ & $13,97$ & $15,07$ & $17,77$ & $21,28$  \\
		\textsc{brazil}                 & $12,50$ & $13,16$ & $14,41$ & $15,63$ & $17,84$ & $21,60$  \\
		\textsc{argentina}        & $12,80$ & $13,18$ & $14,77$ & $16,49$ & $18,03$ & $21,74$  \\
		\textsc{india}                  & $14,10$ & $15,96$ & $16,97$ & $17,56$ & $18,82$ & $22,14$  \\
		\textsc{turkey}             & $32,77$ & $33,72$ & $34,43$ & $35,06$ & $36,63$ & $36,58$  \\
		\textsc{un.arab em.}& $36,71$ & $38,20$ & $39,28$ & $40,43$ & $41,91$ & $42,85$  \\
		\textsc{russia}             & $37,63$ & $38,67$ & $39,50$ & $40,49$ & $42,36$ & $42,99$  \\
		\textsc{kazakistan}   & $38,86$ & $39,77$ & $40,50$ & $41,38$ & $42,67$ & $44,04$  \\
		\textsc{iran}               & $42,71$ & $43,24$ & $43,763$ & $44,18$ & $45,06$ & $44,11$  \\
		\textsc{saudi arabia}& $44,54$ & $45,17$ & $45,55$ & $45,42$ & $45,71$ & $44,61$  \\
		\textsc{china}              & $45,14$ & $45,45$ & $45,91$ & $46,63$ & $47,13$ & $46,57$  \\
		\hline \hline
	\end{tabular}
	\caption{\small Average \textsc{g/t} ratio of Countries over $1000$ runs with different levels of emergency pressure, ranging from $10\%$ to $80\%$. Despite the ranking does not change, differences among different countries are affected and democracies converge to positions close to dictatorships, as shown also in figures.}
	\label{g/t-ranking-emergency}%
\end{table}

\section{Discussion and conclusive remarks}

In the previous sub-sections we tested a modified version of the Hegselmann and Krause (2002) model to analyze how regimes (democracy vs. dictatorship) differently react to emergency pressure. Two different perspectives were investigated: democratic representativeness, i.e. the degree of free circulation of ideas, and time efficiency, i.e. the time necessary to converge to  a final policy decision. Our findings suggest a trade-off between preference representativeness and time delay: the more democratic a country is, the slower the ratio between the global opinion shift per unit of time spent to converge to a final decision (\textsc{g/t}). However, irrespective of regimes, countries tend to converge to similar values of this ratio under emergency pressure. 

This appears to be happened in the case of the  \textsc{covid}-19 pandemic, which we consider here as a typical example of emergency situation, moreover affecting many countries in the world in comparable intervals of time. On January 5 the World Health Organization (\textsc{who}) published its first flagship technical publication contained a risk assessment on the status of patients and the public health response on the cluster of pneumonia cases in Wuhan as reported by China. Two months later, on March 11, the \textsc{who} declared  \textsc{covid}-19 a pandemic, pointing to the over 118,000 cases of the coronavirus illness in over 110 countries and the sustained risk of further global spread. The worry for risk arose containment measures which became gradually more stringent according to emergency pressure. By 26 March, 1.7 billion people worldwide were under some form of lockdown, the more extreme containment measure, which increased to 3.9 billion people by the first week of April. However, the speed with which countries adopted lockdown was different. 

\begin{figure}
	\begin{center}
		\includegraphics[width=4.3in,angle=0]{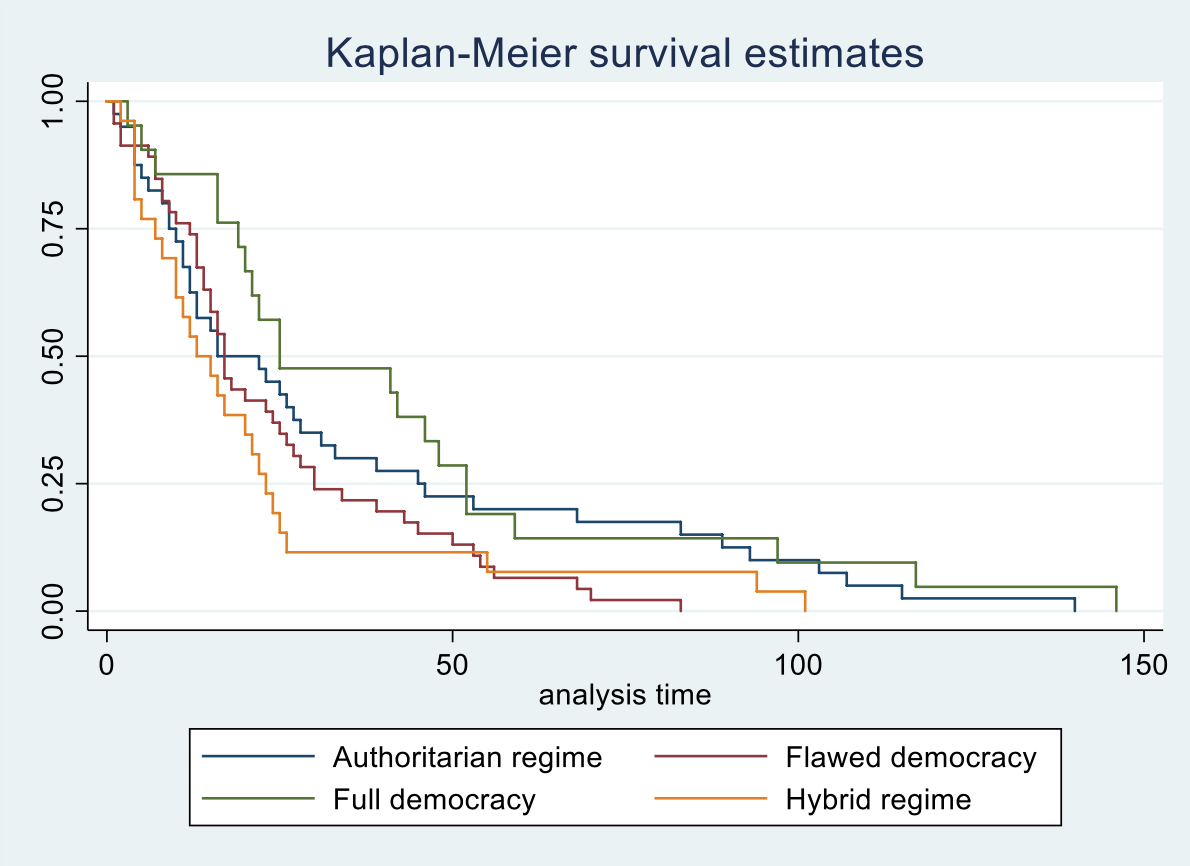}
		\caption{\small The Kaplan-Meier survival estimates for delay response to emergency pressure are plotted for 143 countries, grouped in four regimes following the Economist classification.} 
		\label{kaplan-meier} 
	\end{center}
\end{figure}

Let us assume the time interval between the first confirmed case and lockdown to be a proxy for delay response to emergency pressure. Using data from Oxford  \textsc{covid}-19 Government Response Tracker, edited by the University of Oxford, Figure \ref{kaplan-meier} displays the Kaplan-Meier survival estimates for such a delay for 143 countries. Following the Economist classification, data have been aggregated in four regimes - which can be roughly identified with the four groups of countries recognized in Figure \ref{3d_fig} - and show that plurality of opinions and variety in levels of government may have had a role in determining different temporal replies also in applying emergency measures.

In order to better understand results, it could be useful to split delay time in two sub-period: before and after 50 days. 

It clearly appears that, before 50 days, full democracies are later than other regimes to lockdown, while hybrid regimes, what Cheibub et al. (2010) called civil dictatorships, are quicker to join a given percentage of closure. In particular, considering the negative slope of the first part of the four curves in Figure \ref{kaplan-meier} as proportional to the average  \textsc{g/t} of the corresponding group of countries (under a given emergency pressure level), we can put in relationship their rapidity in adopting the lockdown, and the consequent distortion of preferences, with the average  \textsc{gos} per unit of  \textsc{tdt} found in our simulations (see Table \ref{g/t-ranking-emergency}), which is always smaller for (full or flawed) democracies and higher for (strong or light) dictatorships.

Then, in the long term, regimes collapsed to the same value, but full democracies confirmed to be the longer. This is, again, what we expected from our simulation. In summary, democracy is time-costly, but emergency pressure, as measured by the distance from the first case, forces countries to be quick, taking solutions similar to those adopted by more authoritarian countries.

Such conclusions shed light on the fact that emergencies challenge all institutional configurations and exacerbate limits of their physiologic structure. All regimes exhibit wide differences in the management of critical situations, leading to a quite stable ranking. Our results underline that the political efficiency does not necessarily imply the reduction of the democratic representation: indeed, countries where the social consciousness is high and the free expression of opinions is the normality seem to have some advantage and perform better in our simulations. Nonetheless, in some cases, less efficient democracies experience smaller advantages with respect to authoritarian regimes, which however seem to exhibit a reduction of freedom more proportional than the time reduction gained in terms of decision time. We claim that despite our results cannot be considered as definitive and descriptive of the whole complex picture of political action of Governments, it gives a robust intuition on the existence of both more or less efficient democracies, i.e., more or less politically mature, and more or less enlightened dictatorial regimes. This makes our results realistic and show that the existence of emergencies may unveil fragilities of institutional systems.


\section*{References}

Abelson R.P. (1967). Mathematical Models in Social Psychology. \textit{Advances in Experimental Social Psychology}, vol.3, pp.1-54.

Acemoglu D., Ozdaglar A. (2011). \textit{Opinion Dynamics and Learning in Social Networks}, \textit{Dynamic Games and Applications}, vol.1, pp.3-49.

Adami, C., Schossau, J., Hintze, A. (2016). Evolutionary game theory using agent-based methods. \textit{Physics of Life Reviews}, vol. 19, pp.1-26.

Alon, I., Farrell, M., Li, S. (2020). Regime Type and COVID-19 Response. \textit{FIIB Business Review}, vol.9(3), pp.152–160.

Baekkeskov, E., Rubin, O. (2014). Why pandemic response is unique: powerful experts and hands-off political leaders. \textit{Disaster Prevention and Management}, vol. 23 No. 1, pp. 81-93.

Berengaut, A.A. (2020). Democracies Are Better at Fighting Outbreaks. \textit{The Atlantic}, February, 24, 2020.

Bieber, F. (2020), Authoritarianism in the Time of the Coronavirus, \textit{Foreign Policy}, March 30.

Bonabeau, E. (2002). Agent-based modeling: Methods and techniques for simulating human systems. \textit{PNAS}. 99: 7280-7. 

Buchanan J.M., Tullock G. (1962). \textit{The Calculus of Consent}, The University of Michigan Press, Ann Arbor. 

Bueno de Mesquita B., Smith A., Siverson R.M., Morrow J.D. (2003). \textit{The Logic of Political Survival}. Cambridge, MA: MIT Press

Burkle, F. M. (2020). Declining public health protections within autocratic regimes: impact on global public health security, infectious disease outbreaks, epidemics, and pandemics. \textit{Prehospital and Disaster Medicine}, 35(3), 237-246.


Chakraborti, A., Toke, I. M., Patriarca, M., Abergel, F. (2011a). Econophysics review: I. Empirical facts. \textit{Quantitative Finance}, 11(7), 991-1012.

Chakraborti, A., Toke, I. M., Patriarca, M., Abergel, F. (2011b). Econophysics review: II. Agent-based models. \textit{Quantitative Finance}, 11(7), 1013-1041.

Chang, M. (2011). Agent-based modeling and computational experiments in industrial organization: Growing firms and industries in silico. \textit{Eastern Economic Journal}, vol. 37, 28?34.

Chang, M. (2015). \textit{A computational model of industry dynamics}. Advances in experimental and computable economics book series. London, UK: Routledge.

Dawid, H., Delli Gatti, D. (2018). Agent based macroeconomics. In C. Hommes, B. LeBaron (eds.), \textit{Handbook of Computational Economics} vol. 4, pp. 63?156. Amsterdam, The Netherlands: Elsevier.

Deffuant G., Neau D., Amblard F., and Weisbuch G. (2000) Mixing beliefs among interacting agents. \textit{Advances in Complex Systems} vol. 3, pp. 87-98.

DeGroot M.H. (1974). Reaching a Consensus, \textit{Journal of the American Statistical Association}, vol.69 (345), pp.118-121. doi 10.1080/01621459.1974.10480137

Delli Gatti, D., Gaffeo, E., Gallegati, M., Giulioni, G., Palestrini, A. (2008). \textit{Emergent macroeconomics}. Berlin, Germany: Springer.

Delli Gatti, D., Desiderio, S., Gaffeo, E., Cirillo, P., Gallegati, M. (2011). \textit{Macroeconomics from the bottom-up}. Berlin, Germany: Springer.

Djvadian, S., Chow, J. Y. J. (2017). An agent-based day-to-day adjustment process for modeling ``mobility as a service'' with a two-sided flexible transport market. \textit{Transportation Research Part B}, vol. 104, 36?57.

Epstein J. and Axtell R. (1996).\textit{Growing artificial societies: social science from the bottom}. Brookings Institution Press. pp. 224. ISBN 978-0-262-55025-3.

Fagiolo, G., Roventini, A. (2008). On the scientific status of economic policy: a tale of alternative paradigms (No. 2008/03). \textit{LEM Working Paper Series}.

French J.R.P. jr. (1956) A formal theory of social power. \textit{The Psychological Review}, vol. 63, pp. 181-194.

Friedkin N., Johnsen E. (1999). Social influence networks and opinion change. \textit{Advances in Group Processes}, vol.16, pp.1-29.

Fortunato S. (2005a). On the consensus threshold for the opinion dynamics of Krause-Hegselmann, \textit{International Journal of Modern Physics C}, vol.16 (2), pp. 259-270.

Fortunato S. \textit{et al.} (2005b). Vector opinion dynamics in a bounded confidence consensus model, \textit{International Journal of Modern Physics C}, vol. 16 (10), pp.1535-1551   

Galam S. (2012). \textit{Sociophysics: A Physicist's Modeling of Psycho-Political Phenomena}, Springer.

Green L.V., Kolesar P.J. (2004), Improving Emergency Responsiveness with Management Science, \textit{Management Science}, vol.50 (8), pp.1001-1014.

Harary F. (1959). A criterion for unanimity in French's theory of social power. In Cartwright D. (ed.), \textit{Studies in Social Power}. Ann Arbor, MI, Institute for Social Research, pp. 168-182.

Hegselmann R., Flache A. (1998) Understanding complex social dynamics - a plea for cellular automata based modelling. \textit{Journal of Artificial Societies and Social Simulation}, vol. 1 no. 3. 

Hegselmann R., Krause U., (2002). Opinion Dynamics and Bounded Confidence Models, Analysis and Simulation, \textit{JASSS}, vol.5(3) http://jasss.soc.surrey.ac.uk/5/3/2.html

Hoos I.R. (1971) Information Systems and Public Planning, \textit{Management Science}, vol.17 (10), B-658–B-671. DOI:https://doi.org/10.1287/mnsc.17.10.B658

Khavanag, M.M. (2020),  Authoritarianism, outbreaks, and information politics, \textit{The Lancet}, 5(3), pp.135-136.

Kirman, A. (2011). \textit{Complex economics: Individual and collective rationality}. London, UK: Routledge.

Kleinfeld, R. (2020). Do Authoritarian or Democratic Countries Handle Pandemics Better?. \textit{Commentary} March, 31, 2020.

Krause U. (1997). Soziale Dynamiken mit vielen Interakteuren. Eine Problemskizze. In Krause U. and Stck{\"o}ler M. (eds.) \textit{Modellierung und Simulation von Dynamiken mit vielen interagierenden Akteuren}, Universit{\"a}t Bremen. pp. 37-51.

Krause U. (2000). A discrete nonlinear and non-autonomous model of consensus formation. In Elaydi S., Ladas G., Popenda J. and Rakowski J. (eds.), \textit{Communications in Difference Equations}, Amsterdam: Gordon and Breach Publ. pp. 227-236.

Lehrer K., Wagner C.(1991), \textit{Rational Consensus in Science and Society}, Dordrecht: D.Reidel.

Lorenz J. (2007), Continuous Opinion Dynamics under Bounded Confidence: a Survey, \textit{International Journal of Modern Physics C}, Vol. 18, No. 12, pp. 1819-1838.

Mantegna, R. N., Stanley, H. E. (1999). \textit{Introduction to econophysics: correlations and complexity in finance}. Cambridge University Press.

Morrow, J.D., Bueno de Mesquita, B., Siverson, R., Smith, Al. (2008), Retesting Selectorate Theory: Separating the Effects of W from Other Elements of Democracy, \textit{American Political Science Review}, 102(3), pp.393-400.

Neugart, M., Richiardi, M. (2012). Agent-based models of the labor market. \textit{LABORatorio R. Revelli wp} 125, pp. 164?212

Schmemann, S. (2020). The Virus Comes for Democracy Strongmen think they know the cure for Covid-19. Are they right?. \textit{New York Times}, April, 2.

Schwartz, P. (2012). \textit{The art of the long view: planning for the future in an uncertain world}. John Wiley \& Sons Inc.

Sznajd-Weron K., Sznajd J. (2000), Opinion evolution in closed community, \textit{International Journal Modern Physics C} vol.11 (6) pp.1157-1165.

Tesfatsion, L., Judd, K. L. (Eds.). (2006). Handbook of computational economics: agent-based computational economics. Elsevier.

Tesfatsion, L. (2006). Agent-based computational economics: A constructive approach to economic theory. Handbook of computational economics, 2, pp.831-880.

Von Neuman, J., Burks. (1966). \textit{Theory of self-reproducing automata}. University of Illinois Press. Urbana and London. 

\end{document}